\documentclass{article}
\usepackage{amssymb}
\usepackage{amsfonts}
\usepackage{amsmath}

\usepackage{eurosym}
\usepackage{amssymb}
\usepackage{amsfonts}
\usepackage{amsmath}
\usepackage{graphicx}
\graphicspath{{./Figures/}}
\usepackage[ruled]{algorithm2e}
\usepackage{subcaption}
\usepackage{xcolor}

\usepackage{titlesec}
\usepackage[titletoc,toc,title]{appendix}

\begin{document}

\title{A New Smoothing Technique based on the Parallel Concatenation of
	Forward/Backward Bayesian Filters: Turbo Smoothing}
\author{}
\maketitle

\begin{abstract}
	Recently, a novel method for developing filtering algorithms, based on the
	parallel concatenation of Bayesian filters and called turbo filtering, has
	been proposed. In this manuscript we show how the same conceptual approach
	can be exploited to devise a new smoothing method, called turbo smoothing. A
	turbo smoother combines a turbo filter, employed in its forward pass, with
	the parallel concatenation of two backward information filters used in its
	backward pass. As a specific application of our general theory, a detailed
	derivation of two turbo smoothing algorithms for conditionally linear
	Gaussian systems is illustrated. Numerical results for a specific dynamic
	system evidence that these algorithms can achieve a better
	complexity-accuracy tradeoff than other smoothing techniques recently
	appeared in the literature.
\end{abstract}

\bigskip

\begin{center}
Giorgio M. Vitetta, Pasquale Di Viesti and  Emilio Sirignano 

\vspace{5mm}University of Modena and Reggio Emilia

Department of Engineering "Enzo Ferrari"

Via P. Vivarelli 10/1, 41125 Modena - Italy

email: giorgio.vitetta@unimore.it, pasquale.diviesti2@unibo.it,
emilio.sirignano@unimore.it
\end{center}

\bigskip


\bigskip\vspace{1cm}

\baselineskip0.2 in\newpage

\section{Introduction\label{sec:intro}}

The problem of \emph{Bayesian} \emph{smoothing} for a \emph{state space model%
} (SSM) concerns the development of recursive algorithms able to estimate
the \emph{probability density function} (pdf) of the model state on a given
observation interval, given a batch of noisy measurements acquired over it 
\cite{Anderson_1979}; the estimated pdf is known as a \emph{smoothed} or 
\emph{smoothing} pdf. A general strategy for solving this problem is based
on the so called \emph{two-filter smoothing formula} \cite{Kitagawa_1994}-%
\cite{Bresler_1986}; in fact, this formula allows to compute the required
smoothing density by merging the statistical information generated in the
forward pass of a \emph{Bayesian filtering} method with those evaluated in
the backward pass of a different filtering method, \emph{paired with the
first one} and known as \emph{backward information filtering} (BIF).
Unluckily, closed form solutions for this strategy can be derived for \emph{%
linear Gaussian} and \emph{linear Gaussian mixture} models only \cite%
{Anderson_1979}, \cite{Vo_2012}. For this reason, all the existing smoothing
algorithms based on the above mentioned formula and applicable to general 
\emph{nonlinear} models are approximate and are based on \emph{sequential
Monte Carlo} techniques (e.g., see \cite{Kitagawa_1994}, \cite{Kitagawa_1996}%
, \cite{Fong_2002} and references therein). Unluckily, the adoption of these
algorithms, known as \emph{particle smoothers}, may be hindered by their
complexity, which becomes unmanageable when the dimension of the sample
space for the considered SSM is large.

Recently, a \emph{factor graph} approach has been exploited to devise a new
filtering method, based on the \emph{parallel concatenation} of two (\emph{%
constituent}) Bayesian filters and called \emph{turbo filtering }(TF) \cite%
{Vitetta_2018_TF}. In this manuscript, a new smoothing technique that
employs TF in its \emph{forward} pass and a new BIF scheme, based on the
parallel concatenation of two backward information filters, is developed.
Our derivation of the new BIF method, called \emph{backward information
turbo filtering} (BITF), is based on a \emph{general graphical model}; this
allows us to: a) represent any BITF algorithm as the interconnection of 
\emph{two soft-in soft-out }(SISO) processing modules; b) represent the
iterative processing accomplished by these modules as a message passing
technique; c) derive the expressions of the passed messages by applying the 
\emph{sum-product algorithm} (SPA) \cite{Loeliger_2007}, \cite%
{Kschischang_2001}, together with a specific \emph{scheduling} procedure, to
the graphical model itself; d) show how the statistical information
generated by a BITF algorithm in the backward pass can be merged with those
produced by the paired TF technique in the forward pass in order to evaluate
the required smoothed pdfs. To exemplify the usefulness of this smoothing
method, called \emph{turbo-smoothing} (TS) in the following, we take into
consideration the TF algorithms proposed in \cite{Vitetta_2018_TF} for the
class of \emph{conditionally linear Gaussian} (CLG) SSMs and derive a BITF
algorithm paired with them. This approach leads to the development of two
new TS algorithms, one generating an estimate of the joint smoothing density
over the whole observation interval, the other one an estimate of the
marginal smoothing densities over the same interval. Our computer
simulations for a specific CLG\ SSM evidence that, in the considered case,
the derived TS algorithms perform very closely to the \emph{%
Rao-Blackwellized particle smoothing} (RBPS) technique proposed in \cite%
{Lindsten_2016} and to the particle smoothers devised in \cite{Vitetta_2018}.

The remaining part of this manuscript is organized as follows. A description
of the considered SSMs is illustrated in Section \ref{sec:scenario}. In
Section \ref{sec:Factorgraphs}, a general graphical model on which the
processing accomplished in BITF and TS is based is illustrated; then, a
specific instance of it, referring to a CLG SSM, is developed and the
messages passed over it in BITF are defined. In Section \ref%
{sec:Message-Passing}, the scheduling and the computation of such messages
are described, specific TS algorithms are developed, and the differences and
similarities between these algorithms and other smoothing techniques are
briefly analysed. A comparison, in terms of accuracy and execution time,
between the proposed techniques and three smoothers recently appeared in the
literature is provided in Section \ref{num_results} for a specific CLG SSM.
Finally, some conclusions are offered in Section \ref{sec:conc}.

\emph{Notations}: The same notations as refs. \cite{Vitetta_2018}, \cite%
{Vitetta_2018_TF} and \cite{Vitetta_2019} are adopted.

\section{Model Description\label{sec:scenario}}

In this manuscript we focus on a discrete-time SSM whose $D$-dimensional 
\emph{hidden state} in the $l$-th interval is denoted $\mathbf{x}%
_{l}\triangleq \lbrack x_{0,l},x_{1,l},...,$ $x_{D-1,l}]^{T}$, and whose 
\emph{state update} and \emph{measurement models} are expressed by 
\begin{equation}
\mathbf{x}_{l+1}=\mathbf{f}_{l}\left( \mathbf{x}_{l}\right) +\mathbf{w}_{l}
\label{eq:X_update}
\end{equation}%
and%
\begin{eqnarray}
\mathbf{y}_{l} &\triangleq &[y_{0,l},y_{1,l},...,y_{P-1,l}]^{T}  \notag \\
&=&\mathbf{h}_{l}\left( \mathbf{x}_{l}\right) +\mathbf{e}_{l},
\label{meas_mod}
\end{eqnarray}%
respectively. Here, $\mathbf{f}_{l}\left( \mathbf{x}_{l}\right) $ ($\mathbf{h%
}_{l}\left( \mathbf{x}_{l}\right) $) is a time-varying $D$-dimensional ($P$%
-dimensional) real function and $\mathbf{w}_{l}$ ($\mathbf{e}_{l}$) the $l$%
-th element of the process (measurement) noise sequence $\left\{ \mathbf{w}%
_{k}\right\} $ ($\left\{ \mathbf{e}_{k}\right\} $); this sequence consists
of $D$-dimensional ($P$-dimensional) \emph{independent and identically
distributed} (iid) Gaussian noise vectors, each characterized by a zero mean
and a covariance matrix $\mathbf{C}_{w}$ ($\mathbf{C}_{e}$). Moreover,
statistical independence between $\left\{ \mathbf{e}_{k}\right\} $ and $\{%
\mathbf{w}_{k}\}$ is assumed.

In the following, two additional mathematical representations for the
considered SSM are also exploited. The first one is approximate, being
employed by an \emph{extended Kalman filter} (EKF); in fact, it is based on
the \emph{linearized} versions of eqs. (\ref{eq:X_update}) and (\ref%
{meas_mod}), namely (e.g., see \cite[pp. 194-195]{Anderson_1979})%
\begin{equation}
\mathbf{x}_{l+1}=\mathbf{F}_{l}\mathbf{x}_{l}+\mathbf{u}_{l}+\mathbf{w}_{l}
\label{state_up_approx}
\end{equation}%
and%
\begin{equation}
\mathbf{y}_{l}=\mathbf{H}_{l}^{T}\mathbf{x}_{l}+\mathbf{v}_{l}+\mathbf{e}%
_{l},  \label{meas_mod_approx}
\end{equation}%
respectively; here, $\mathbf{F}_{l}\triangleq \lbrack \partial \mathbf{f}%
_{l}\left( \mathbf{x}\right) /\partial \mathbf{x}]_{\mathbf{x=x}_{fe,l}}$, $%
\mathbf{x}_{fe,l}$ is the (forward) \emph{estimate} of $\mathbf{x}_{l}$
evaluated by the EKF in its $l$-th recursion, $\mathbf{u}_{l}\triangleq 
\mathbf{f}_{l}\left( \mathbf{x}_{fe,l}\right) -\mathbf{F}_{l}\mathbf{x}%
_{fe,l}$, $\mathbf{H}_{l}^{T}\triangleq \lbrack \partial \mathbf{h}%
_{l}\left( \mathbf{x}\right) /\partial \mathbf{x}]_{\mathbf{x=x}_{fp,l}}$, $%
\mathbf{x}_{fp,l}$ is the (forward) \emph{prediction} $\mathbf{x}_{l}$
computed by the EKF in its $(l-1)$-th recursion and $\mathbf{v}%
_{l}\triangleq \mathbf{h}_{l}\left( \mathbf{x}_{fp,l}\right) -\mathbf{H}%
_{l}^{T}\mathbf{x}_{fp,l}$.

The second representation is based on the additional assumption that the SSM
described by eqs. (\ref{eq:X_update})-(\ref{meas_mod}) is CLG \cite%
{Lindsten_2016}, \cite{Schon_2005}, so that its state vector in the $l$-th
interval can be partitioned as $\mathbf{x}_{l}=[(\mathbf{x}_{l}^{(L)})^{T},(%
\mathbf{x}_{l}^{(N)})^{T}]^{T}$; here, $\mathbf{x}_{l}^{(L)}\triangleq
\lbrack x_{0,l}^{(L)}$, $x_{1,l}^{(L)},...,x_{D_{L}-1,l}^{(L)}]^{T}$ ($%
\mathbf{x}_{l}^{(N)}\triangleq \lbrack
x_{0,l}^{(N)},x_{1,l}^{(N)},...,x_{D_{N}-1,l}^{(N)}]^{T}$) is the so called 
\emph{linear }(\emph{nonlinear}) \emph{component} of $\mathbf{x}_{l}$, with $%
D_{L}<D$ ($D_{N}=D-D_{L}$). For this reason, following \cite{Vitetta_2018}, 
\cite{Vitetta_2019} and \cite{Schon_2005}, the models 
\begin{equation}
\mathbf{x}_{l+1}^{(Z)}=\mathbf{A}_{l}^{(Z)}\left( \mathbf{x}%
_{l}^{(N)}\right) \mathbf{x}_{l}^{(L)}+\mathbf{f}_{l}^{(Z)}\left( \mathbf{x}%
_{l}^{(N)}\right) +\mathbf{w}_{l}^{(Z)}  \label{eq:XL_update}
\end{equation}%
and%
\begin{equation}
\mathbf{y}_{l}=\mathbf{g}_{l}\left( \mathbf{x}_{l}^{(N)}\right) +\mathbf{B}%
_{l}\left( \mathbf{x}_{l}^{(N)}\right) \mathbf{x}_{l}^{(L)}+\mathbf{e}_{l}
\label{eq:y_t}
\end{equation}%
are adopted for the update of the \emph{linear} ($Z=L$) and \emph{nonlinear}
($Z=N$) components and for the measurement vector, respectively. In the
state update model (\ref{eq:XL_update}), $\mathbf{f}_{l}^{(Z)}(\mathbf{x}%
_{l}^{(N)})$ ($\mathbf{A}_{l}^{(Z)}(\mathbf{x}_{l}^{(N)})$) is a
time-varying $D_{Z}$-dimensional real function ($D_{Z}\times D_{L}$ real
matrix) and $\mathbf{w}_{l}^{(Z)}$ consists of the first $D_{L}$ (last $%
D_{N} $) elements of $\mathbf{w}_{l}$ if $Z=L$ ($Z=N$); independence between 
$\{\mathbf{w}_{k}^{(L)}\}$ and $\{\mathbf{w}_{k}^{(N)}\}$ is also assumed
for simplicity and the covariance matrix $\mathbf{w}_{k}^{(L)}$ ($\mathbf{w}%
_{k}^{(N)}$) is denoted $\mathbf{C}_{w}^{(L)}$ ($\mathbf{C}_{w}^{(N)}$). In
the measurement model (\ref{eq:y_t}), instead, $\mathbf{g}_{l}(\mathbf{x}%
_{l}^{(N)})$ ($\mathbf{B}_{l}(\mathbf{x}_{l}^{(N)})$) is a time-varying $P$%
-dimensional real function ($P\times D_{L}$ real matrix).

In the next two Sections we focus on the problem of developing algorithms
for the estimation of a) the \emph{joint} \emph{smoothed pdf} $f(\mathbf{x}%
_{1:T}|\mathbf{y}_{1:T})$ (problem \textbf{P.1}) and b) the sequence of 
\emph{marginal} \emph{smoothed pdfs\ }$\{f(\mathbf{x}_{l}|\mathbf{y}%
_{1:T}),\,l=1,2,...,T\}$ (problem \textbf{P.2}); here, $T$ is the duration
of the observation interval and $\mathbf{y}_{1:T}=\left[ \mathbf{y}_{1}^{T},%
\mathbf{y}_{2}^{T},...,\mathbf{y}_{T}^{T}\right] ^{T}$ is a $T\cdot P$%
-dimensional vector. It is important to point out that: a) in solving both
problems P.1 and P.2, the prior knowledge of the pdf $f(\mathbf{x}_{1})$ of
the initial state is assumed; b) in principle, if an estimate of the joint
pdf $f(\mathbf{x}_{1:T}|\mathbf{y}_{1:T})$ is available, estimates of all
the posterior $\{f(\mathbf{x}_{l}|\mathbf{y}_{1:T})\}$ can be evaluated by 
\emph{marginalization}.

\section{Graphical Modelling for the Parallel Concatenation of Bayesian
Information Filters \label{sec:Factorgraphs}}

In this Section, we derive the graphical models on which BITF and TS
techniques are based. More specifically, starting from the factor graph
representing Bayesian smoothing \cite{Vitetta_2018}, we first develop a
general graphical model for the parallel concatenation of two backward
information filters. Then, a specific instance of this model is devised for
the case in which the forward filters are an EKF and a \emph{particle filter}
(PF), and the considered SSM is CLG.

\subsection{Graphical Model for the Parallel Concatenation of Bayesian
Information Filters and Message Passing over it\label{General_Models}}

\begin{figure}[tbp]
\centering
\includegraphics[width=0.70\textwidth]{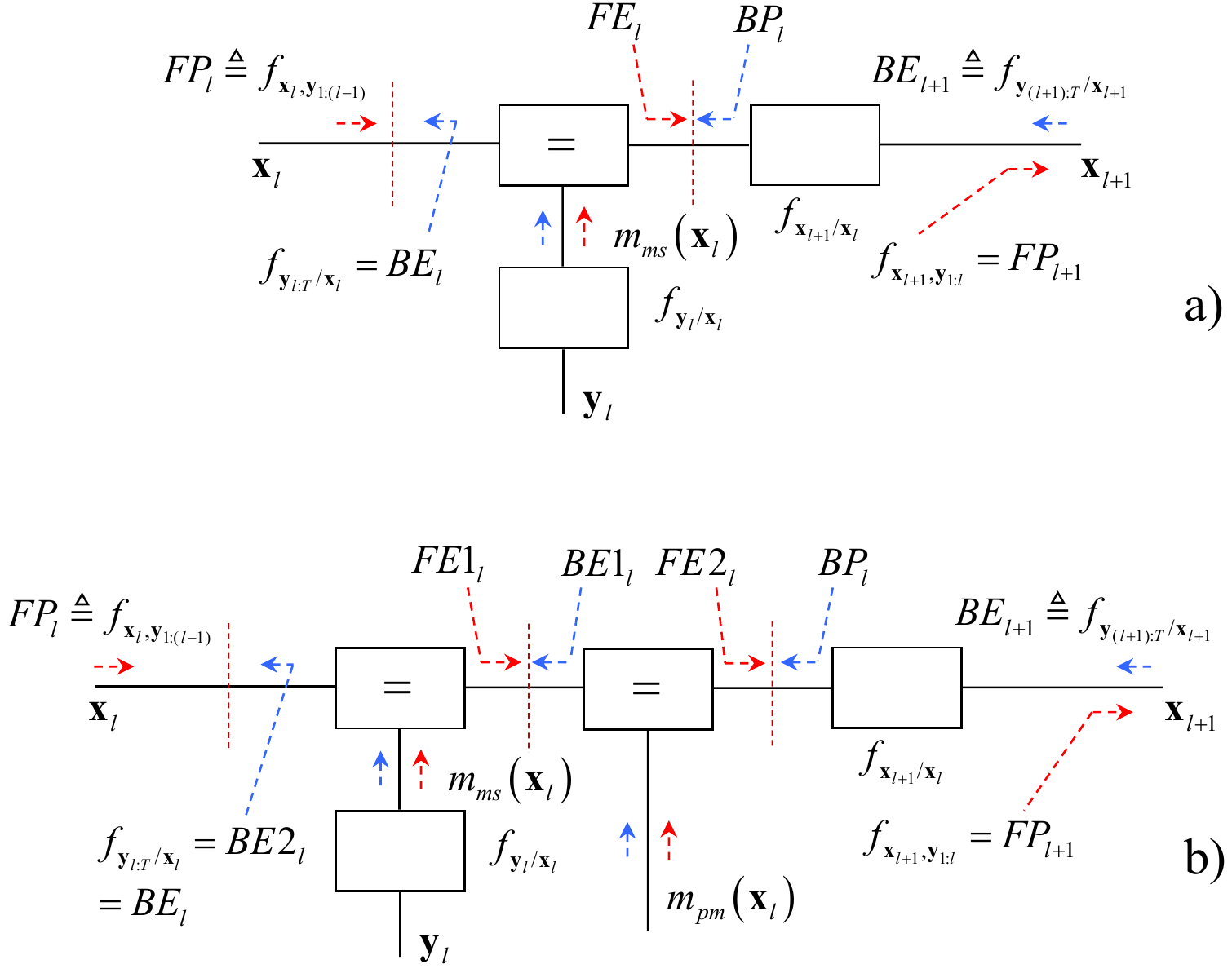}
\caption{Message passing in Bayesian filtering and BIF in the cases of: a)
availability of measurements only; b) availability of measurements and
pseudo-measurements. The flow of messages in the forward (backward) pass are
indicated by red (blue) arrows, respectively; the brown vertical lines
cutting each graph identify the partitioning associated with a) formulas (%
\protect\ref{factorisation1}) (left cut) and (\protect\ref{factorisation2})
(right cut) and b) formulas (\protect\ref{factorisation3a}) (left cut), (%
\protect\ref{factorisation3b}) (central cut) and (\protect\ref%
{factorisation3c}) (right cut)}
\label{Fig_1}
\end{figure}

The development of our BIF algorithms is based on the graphical approach
illustrated in ref. \cite[Sec. III]{Vitetta_2018}. This approach consists in
representing \emph{Bayesian filtering} and BIF as two recursive algorithms
that compute, on the basis of the SPA, a set of probabilistic messages
passed along the \emph{same} (cycle free) factor graph; this graph is
illustrated Fig. \ref{Fig_1}-a) and refers to a SSM\ characterized by the
Markov model $f(\mathbf{x}_{l+1}|\mathbf{x}_{l})$ and the measurement model $%
f(\mathbf{y}_{l}|\mathbf{x}_{l})$. More specifically, in the $l$-th
recursion of Bayesian filtering\emph{, }messages are passed along the
considered graph in the \emph{forward} direction; moreover, the messages $%
\vec{m}_{fe}\left( \mathbf{x}_{l}\right) =f\left( \mathbf{x}_{l},\mathbf{y}%
_{1:l}\right) $ and $\vec{m}_{fp}\left( \mathbf{x}_{l+1}\right) =f(\mathbf{x}%
_{l+1},\mathbf{y}_{1:l})$ (denoted $FE_{l}$ and $FP_{l+1}$, respectively, in
Fig. \ref{Fig_1} and conveying a \emph{forward estimate} of $\mathbf{x}_{l}$
and a \emph{forward prediction} of $\mathbf{x}_{l+1}$, respectively) are
computed on the basis of the input message 
\begin{equation}
\vec{m}_{fp}\left( \mathbf{x}_{l}\right) \triangleq f\left( \mathbf{x}_{l},%
\mathbf{y}_{1:(l-1)}\right) ,  \label{m_fp_l_SSM}
\end{equation}%
for $l=2$, $3$, $...$, $T$ ($\vec{m}_{fp}\left( \mathbf{x}_{1}\right) =f(%
\mathbf{x}_{1})$). Dually, in the $(T-l)$-th recursion of BIF, messages are
passed along the considered graph in the \emph{backward} direction, and the
messages $\overset{\leftarrow }{m}_{bp}\left( \mathbf{x}_{l}\right) =f(%
\mathbf{y}_{(l+1):T}|\mathbf{x}_{l})$ and $\overset{\leftarrow }{m}%
_{be}\left( \mathbf{x}_{l}\right) =f(\mathbf{y}_{l:T}|\mathbf{x}_{l})$
(denoted $BP_{l}$ and $BE_{l}$, respectively, in Fig. \ref{Fig_1} and
conveying a \emph{backward prediction} of $\mathbf{x}_{l}$ and a \emph{%
backward estimate} of $\mathbf{x}_{l+1}$, respectively) are computed on the
basis of the input message%
\begin{equation}
\overset{\leftarrow }{m}_{be}\left( \mathbf{x}_{l+1}\right) \triangleq
f\left( \mathbf{y}_{(l+1):T}\left\vert \mathbf{x}_{l+1}\right. \right) ,
\label{m_be_l+1_SSM}
\end{equation}%
with $l=T-2,T-3,...,1$ ($\overset{\leftarrow }{m}_{be}\left( \mathbf{x}%
_{T}\right) =f\left( \mathbf{y}_{T}|\mathbf{x}_{T}\right) $). Once the
backward pass is over, a solution to problem \textbf{P.2} becomes available,
since the marginal smoothed pdf $f\left( \mathbf{x}_{l},\mathbf{y}%
_{1:T}\right) $ can be evaluated as\footnote{%
Note that, similarly as \cite{Vitetta_2018} and \cite{Vitetta_2019}, the 
\emph{joint} pdf $f(\mathbf{x}_{l},\mathbf{y}_{1:T})$ is considered here in
place of the \emph{posterior} pdf $f(\mathbf{x}_{l}|\mathbf{y}_{1:T})$.}%
\begin{equation}
f\left( \mathbf{x}_{l},\mathbf{y}_{1:T}\right) =\vec{m}_{fp}\left( \mathbf{x}%
_{l}\right) \overset{\leftarrow }{m}_{be}\left( \mathbf{x}_{l}\right) \,
\label{factorisation1}
\end{equation}%
or, equivalently, as 
\begin{equation}
f\left( \mathbf{x}_{l},\mathbf{y}_{1:T}\right) =\vec{m}_{fe}\left( \mathbf{x}%
_{l}\right) \overset{\leftarrow }{m}_{bp}\left( \mathbf{x}_{l}\right) ,
\label{factorisation2}
\end{equation}%
with $l=1,2,...,T$. Note that, from a graphical viewpoint, formulas (\ref%
{factorisation1}) and (\ref{factorisation2}) can be related with the two
different partitionings of the graph shown in Fig. \ref{Fig_1}-a) (where a
specific partitioning is identified by a brown dashed vertical line cutting
the graph in two parts).

In ref. \cite{Vitetta_2018_TF}\ it has been also shown that the factor
graph\ illustrated in Fig. \ref{Fig_1}-a) can be employed as a building
block in the development of a larger graphical model that represents a \emph{%
turbo filtering} scheme, i.e. the \emph{parallel concatenation} of two (%
\emph{constituent}) Bayesian filters (denoted F$_{1}$ and F$_{2}$ in the
following). In this model, the graphs referring to F$_{1}$ and F$_{2}$ are
interconnected in order to allow the mutual exchange of statistical
information in the form of \emph{pseudo-measurements} (conveyed by
probabilistic messages). From a graphical viewpoint, the exploitation of
these additional information in each filter requires:

a) modifying the graph shown in Fig. \ref{Fig_1}-a) in a way that each
constituent filter can benefit from the pseudo-measurements provided by the
other filter through an additional \emph{measurement update};

b) developing message passing algorithms over a proper graphical model for
1) the \emph{conversion} of the statistical information generated by each\
constituent filter into a form useful to the other one and 2) the \emph{%
generation}, inside each constituent filter, of the statistical information
to be made available to the other filter.

As far as the need expressed at point a) is concerned, the graph of Fig. \ref%
{Fig_1}-a) can be easily modified by adding a new equality node and a new
edge along which the message $m_{pm}\left( \mathbf{x}_{l}\right) $,
conveying pseudo-measurement information, is passed; this results in the
factor graph shown in Fig. \ref{Fig_1}-b). Note that, in the new graphical
model, two \emph{forward estimates} (\emph{backward estimates}) are computed
in the \emph{forward} (\emph{backward}) pass. The first estimate,
represented by $\vec{m}_{fe1}\left( \mathbf{x}_{l}\right) $ ($\overset{%
\leftarrow }{m}_{be1}\left( \mathbf{x}_{l}\right) $) is generated by merging 
$\vec{m}_{fp}\left( \mathbf{x}_{l}\right) $ ($\overset{\leftarrow }{m}%
_{bp}\left( \mathbf{x}_{l}\right) $) with the message $m_{ms}\left( \mathbf{x%
}_{l}\right) $ ($m_{pm}\left( \mathbf{x}_{l}\right) $) conveying measurement
(pseudo-measurement) information, whereas the second one, represented by $%
\vec{m}_{fe2}\left( \mathbf{x}_{l}\right) $ ($\overset{\leftarrow }{m}%
_{be2}\left( \mathbf{x}_{l}\right) =\overset{\leftarrow }{m}_{be}\left( 
\mathbf{x}_{l}\right) $), is evaluated by merging $\vec{m}_{fe1}\left( 
\mathbf{x}_{l}\right) $ ($\overset{\leftarrow }{m}_{be1}\left( \mathbf{x}%
_{l}\right) $) with the message $m_{pm}\left( \mathbf{x}_{l}\right) $ ($%
m_{ms}\left( \mathbf{x}_{l}\right) $). Moreover, similarly as the previous
case, the smoothed pdf $f\left( \mathbf{x}_{l},\mathbf{y}_{1:T}\right) $ can
be computed as 
\begin{eqnarray}
f\left( \mathbf{x}_{l},\mathbf{y}_{1:T}\right) &=&\vec{m}_{fp}\left( \mathbf{%
x}_{l}\right) \overset{\leftarrow }{m}_{be2}\left( \mathbf{x}_{l}\right)
\label{factorisation3a} \\
&=&\vec{m}_{fe1}\left( \mathbf{x}_{l}\right) \overset{\leftarrow }{m}%
_{be1}\left( \mathbf{x}_{l}\right)  \label{factorisation3b} \\
&=&\vec{m}_{fe2}\left( \mathbf{x}_{l}\right) \overset{\leftarrow }{m}%
_{bp}\left( \mathbf{x}_{l}\right) ;  \label{factorisation3c}
\end{eqnarray}%
note also that each of these factorisations can be associated with one of
the three distinct vertical cuts drawn in Fig. \ref{Fig_1}-b).

As far as point b) is concerned, in\ ref. \cite{Vitetta_2018_TF} it is shown
that, in any TF scheme, all the processing tasks related to the conversion
(generation) of the statistical information emerging from (feeding) each
constituent filter can be easily incorporated in a single\emph{\ module},
called \emph{soft-in soft-out} (SISO) module and whose overall processing
can be represented as message passing over a graphical model including the
factor graph shown in Fig. \ref{Fig_1}-b). For this reason, any TF scheme
can be devised by linking (i.e., by concatenating) two SISO modules, each
incorporating a specific filtering algorithm and exchanging probabilistic
information in an iterative fashion. It is also important to point out that
the two constituent filters are not required to estimate the whole system
state. For this reason, in the following, we assume that: a) the filter F$%
_{i}$ estimates the portion $\mathbf{x}_{l}^{(i)}$ (with $i=1$ and $2$) of
the state vector $\mathbf{x}_{l}$ (the size of $\mathbf{x}_{l}^{(i)}$ is
denoted $D_{i}$, with $D_{i}\leq D$); b) the portion of $\mathbf{x}_{l}$ not
included in $\mathbf{x}_{l}^{(i)}$ is denoted $\mathbf{\bar{x}}_{l}^{(i)}$,
so that the equalities $\mathbf{x}_{l}=[(\mathbf{x}_{l}^{(1)})^{T},(\mathbf{%
\bar{x}}_{l}^{(1)})^{T}]^{T}$ or $\mathbf{x}_{l}=[(\mathbf{\bar{x}}%
_{l}^{(2)})^{T},(\mathbf{x}_{l}^{(2)})^{T}]^{T}$ hold. However, the vector $%
\mathbf{\bar{x}}_{l}^{(1)}$ ($\mathbf{\bar{x}}_{l}^{(2)}$) is required to be
part of (or, at most, to coincide with) $\mathbf{x}_{l}^{(2)}$ ($\mathbf{x}%
_{l}^{(1)}$), so that an \emph{overall estimate} of the system state $%
\mathbf{x}_{l}$ can be always generated on the basis of the posterior pdfs
of $\mathbf{x}_{l}^{(1)}$\ and\ $\mathbf{x}_{l}^{(2)}$ evaluated by F$_{1}$
and F$_{2}$, respectively. In fact, this constraint on $\mathbf{\bar{x}}%
_{l}^{(1)}$ and $\mathbf{\bar{x}}_{l}^{(2)}$ leads to the conclusion that,
generally speaking, the portion $\mathbf{x}%
_{l}^{(12)}=[x_{D-D_{2},l},x_{D-D_{2}+1,l},...,x_{D_{1}-1,l}]^{T}$ of $%
\mathbf{x}_{l}$, collecting $N_{d}\triangleq D_{1}+D_{2}-D$ state variables,
is estimated by both F$_{1}$ and F$_{2}$, being shared by $\mathbf{x}%
_{l}^{(1)}$\ and\ $\mathbf{x}_{l}^{(2)}$.

A similar conceptual approach is followed in the remaining part of this
Paragraph to derive the general representation of the BIF technique paired
with a given TF scheme, that is, briefly, a \emph{backward information turbo
filtering} (BITF) technique. This means that:

1) The general architecture we propose for BITF is based on the parallel
concatenation of two \emph{constituent} Bayesian information filters, that
are denoted BIF$_{1}$ and BIF$_{2}$ in the following.

2) The processing accomplished by BIF$_{1}$ (BIF$_{2}$) is represented as a
message passing algorithm over the \emph{same} graphical model as F$_{1}$ (F$%
_{2}$).

3) BITF processing can be represented as the iterative exchange of
probabilistic information between two distinct SISO modules.

4) The $i$-th SISO module (with $i=1$ and $2$) incorporates a specific BIF
algorithm, that can be represented as a message passing over a factor graph
similar to that shown in Fig. \ref{Fig_1}-b) and that estimates the portion $%
\mathbf{x}_{l}^{(i)}$ of $\mathbf{x}_{l}$.

\begin{figure}[tbp]
\centering
\includegraphics[width=0.70\textwidth]{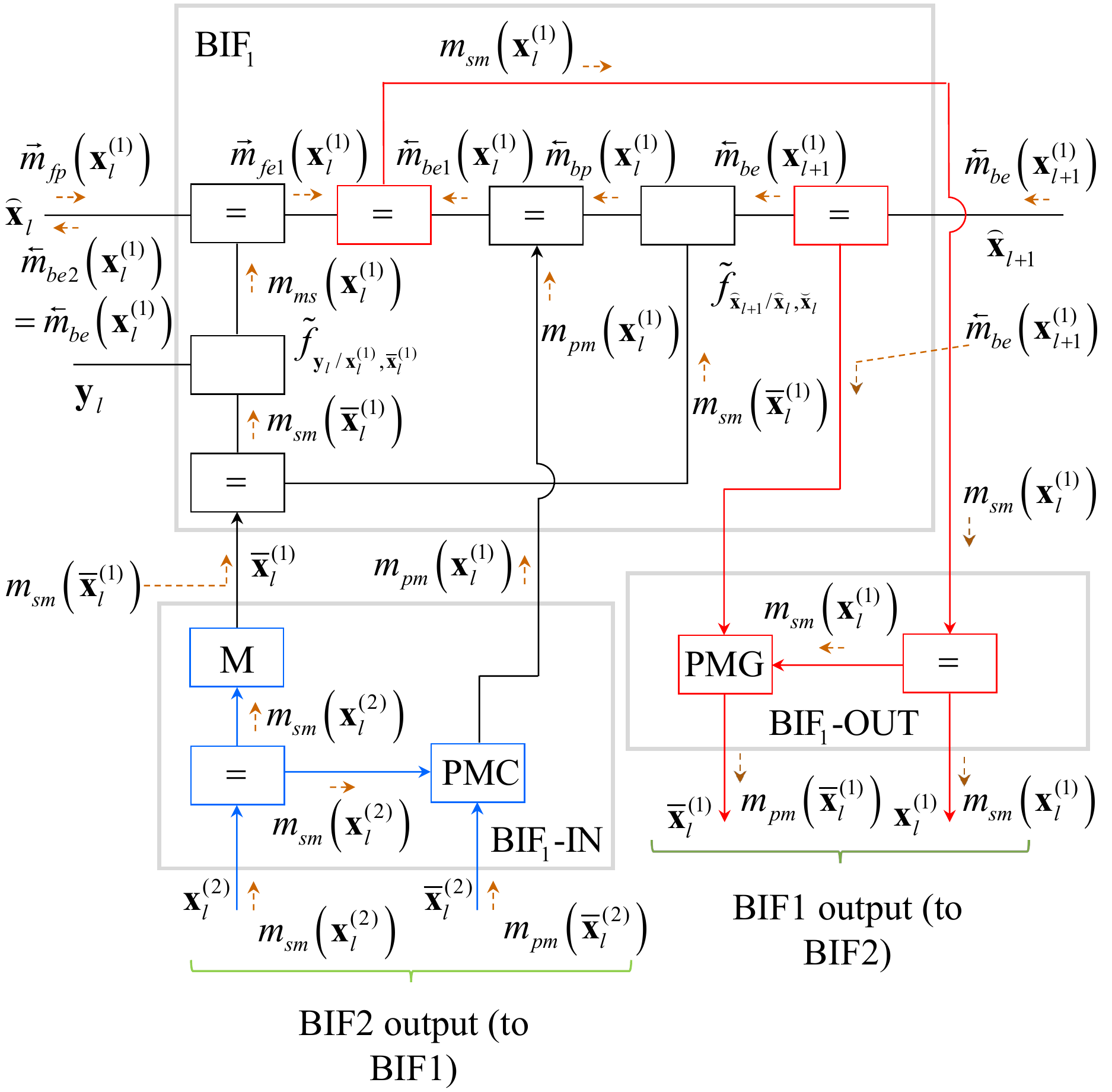}
\caption{Graphical model representing the processing accomplished by the
proposed SISO module based on BIF$_{1}$. Black and blue (red) lines are used
to identify the edges and the blocks related to backward filtering and
processing of information coming from BIF$_{2}$ (made available to BIF$_{2}$%
), respectively.}
\label{Fig_2}
\end{figure}

The graphical model developed for the SISO module based on BIF$_{1}$ is
shown in Fig. \ref{Fig_2}. In this Figure, to ease the interpretation of
message passing, three rectangles, labeled as BIF$_{1}$-IN, BIF$_{1}$ and BIF%
$_{1}$-OUT, have been drawn; this allow us to easily identify the portions
of the graphical model involved in a) the \emph{conversion}\ of the
statistical information provided from BIF$_{2}$ into a form useful to BIF$%
_{1}$, b) BIF$_{1}$ processing and c) the \emph{generation} of the
statistical information made available by BIF$_{1}$ to BIF$_{2}$,
respectively. \ A detailed description of the signal processing tasks
accomplished within\ each portion is provided below.

BIF$_{1}$-IN - The statistical information provided by BIF$_{2}$ to the
considered SISO module is condensed in the messages $m_{sm}(\mathbf{x}%
_{l}^{(2)})$ and $m_{pm}(\mathbf{\bar{x}}_{l}^{(2)})$; these convey a
smoothed estimate of $\mathbf{x}_{l}^{(2)}$ and pseudo-measurement
information about $\mathbf{\bar{x}}_{l}^{(2)}$, respectively. The first
message is processed in two different ways. In fact, on the one hand, it is
marginalised in the block labelled by the letter M (see Fig. \ref{Fig_2}) in
order to generate the pdf $m_{sm}(\mathbf{\bar{x}}_{l}^{(1)})$ (do not
forget that the state vector $\mathbf{\bar{x}}_{l}^{(1)}$ is included in $%
\mathbf{x}_{l}^{(2)}$); on the other hand, $m_{sm}(\mathbf{x}_{l}^{(2)})$ is
processed jointly with $m_{pm}(\mathbf{\bar{x}}_{l}^{(2)})$ in order to
generate the message $m_{pm}(\mathbf{x}_{l}^{(1)})$ conveying
pseudo-measurement information about $\mathbf{x}_{l}^{(1)}$ (this is
accomplished in the block called PM\emph{\ conversion}, PMC; see Fig. \ref%
{Fig_2}). Then, the messages $m_{sm}(\mathbf{\bar{x}}_{l}^{(1)})$ and $%
m_{pm}(\mathbf{x}_{l}^{(1)})$ are passed to BIF$_{1}$.

BIF$_{1}$ - The message passing accomplished in this part refers to the BIF
algorithm paired with F$_{1}$. The graphical model developed for it and the
message passing accomplished over it are based on Fig. \ref{Fig_1}-b). Note,
however, that: a) the message passing aims at computing the (backward)
predicted density $\overset{\leftarrow }{m}_{bp}(\mathbf{x}_{l}^{(1)})$ and
the (backward) filtered density $\overset{\leftarrow }{m}_{be2}(\mathbf{x}%
_{l}^{(1)})=\overset{\leftarrow }{m}_{be}(\mathbf{x}_{l}^{(1)})$ and on the
basis of the backward estimate $\overset{\leftarrow }{m}_{be}(\mathbf{x}%
_{l+1}^{(1)})$ originating from the previous recursion, and of the messages $%
m_{sm}(\mathbf{\bar{x}}_{l}^{(1)})$ and $m_{pm}(\mathbf{x}_{l}^{(1)})$
provided by BIF$_{1}$-IN; b) an approximate model of the considered SSM
could be adopted in the evaluation of these densities. For this reason,
generally speaking, we can assume that the BIF$_{1}$ algorithm is based on
the \emph{Markov model} $\tilde{f}(\mathbf{x}_{l+1}^{(1)}|\mathbf{x}%
_{l}^{(1)},\mathbf{\bar{x}}_{l}^{(1)})$ and on the \emph{observation model} $%
\tilde{f}(\mathbf{y}_{l}|\mathbf{x}_{l}^{(1)},\mathbf{\bar{x}}_{l}^{(1)})$,
representing the \emph{exact} models $f(\mathbf{x}_{l+1}^{(1)}|\mathbf{x}%
_{l}^{(1)},\mathbf{\bar{x}}_{l}^{(1)})$ and $f(\mathbf{y}_{l}|\mathbf{x}%
_{l}^{(1)},\mathbf{\bar{x}}_{l}^{(1)})$, respectively, or \emph{%
approximations} of one or both of them. Note also that, in both the second
measurement update and the time update accomplished by this algorithm,
marginalization with respect to the unknown state component $\mathbf{\bar{x}}%
_{l}^{(1)}$ is made possible by the availability of the message $m_{sm}(%
\mathbf{\bar{x}}_{l}^{(1)})$.

BIF$_{1}$-OUT - This part is fed by the backward estimate $\overset{%
\leftarrow }{m}_{be}(\mathbf{x}_{l+1}^{(1)})$ of $\mathbf{x}_{l+1}^{(1)}$
and by the smoothed estimate $m_{sm}(\mathbf{x}_{l}^{(1)})$ of $\mathbf{x}%
_{l}^{(1)}$ (available after that the first measurement update has been
accomplished by F$_{1}$). The second message follows two different paths,
since a) it is passed to the other SISO module as it is and b) it is jointly
processed with $\overset{\leftarrow }{m}_{be}(\mathbf{x}_{l+1}^{(1)})$ in
order to generate the pseudo-measurement message $m_{pm}(\mathbf{\bar{x}}%
_{l}^{(1)})$ feeding the other SISO module; the last task is accomplished in
the \emph{pseudo-measurement\ generation} (PMG) block.

A graphical model structurally identical to the one shown in Fig. \ref{Fig_2}
can be easily drawn for the SISO module based on BIF$_{2}$ by interchanging $%
\mathbf{x}_{l}^{(1)}$ ($\mathbf{\bar{x}}_{l}^{(1)}$) with $\mathbf{x}%
_{l}^{(2)}$ ($\mathbf{\bar{x}}_{l}^{(2)}$). Merging the graphical model
shown in Fig. \ref{Fig_2} with its counterpart referring to BIF$_{2}$
results in the parallel concatenation architecture illustrated in Fig. \ref%
{Fig_3} (details about the underlying graphical model are omitted for
simplicity) and on which TS is based. It is important to point out that:

1. The overall graphical model derived for TS, unlike the one illustrated in
Fig. \ref{Fig_1}, is not cycle free; therefore, the application of the SPA
to it requires defining a proper message scheduling and, generally speaking,
results in iterative algorithms.

2. At the end of the $l$-th recursion of a TS algorithm, two smoothed
densities, namely $m_{sm}(\mathbf{x}_{l}^{(1)})$ and $m_{sm}(\mathbf{x}%
_{l}^{(2)})$, are available. This raises the problem of how these
statistical information can be fused in order to get a single pdf for (and,
in particular, a single smoothed estimate of) the $N_{d}$-dimensional
portion $\mathbf{x}_{l}^{(12)}$ of $\mathbf{x}_{l}$ estimated by both F$_{1}$%
/BIF$_{1}$ and F$_{2}$/BIF$_{2}$. Unluckily, this remains an open issue. In
our computer simulations, a simple selection strategy has been adopted in
state estimation, since one of the two smoothed estimates of $\mathbf{x}%
_{l}^{(12)}$ has been systematically discarded.

\begin{figure}[tbp]
\centering
\includegraphics[width=0.70\textwidth]{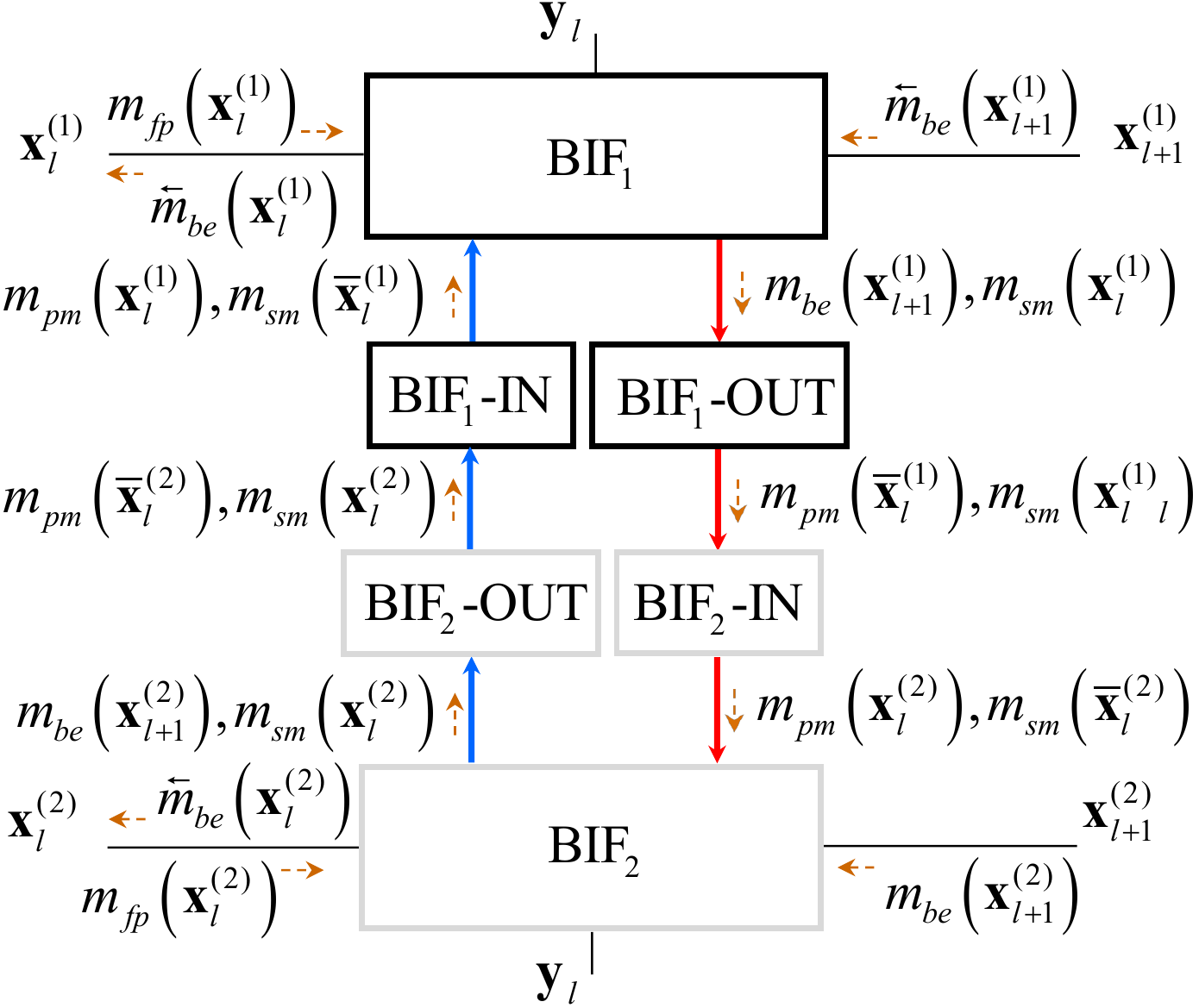}
\caption{Parallel concatenation of two SISO modules based on distinct
backward information filters (denoted BIF$_{1}$ and BIF$_{2}$); the flow of
the messages exchanged between them is indicated by brown arrows.}
\label{Fig_3}
\end{figure}

\subsection{A Graphical Model for the Parallel Concatenation of the Bayesian
Information Filters Paired with an Extended Kalman Filter and a Particle
Filter\label{Graph_mod_CLG}}

In the remaining part of this manuscript we focus on a specific instance of
the proposed TS architecture, since we make the same specific choices as 
\cite{Vitetta_2018_TF}\ for both the SSM and the filters employed in the
forward pass. In particular, we focus on the CLG\ SSM described in Section %
\ref{sec:scenario} and assume that:

1) BIF$_{1}$ is the backward filter associated with an EKF operating over
the whole system state (so that $\mathbf{x}_{l}^{(1)}=\mathbf{x}_{l}$ and $%
\mathbf{\bar{x}}_{l}^{(1)}$ is empty). In other words, BIF$_{1}$ is a
backward Kalman filter based on a linearised model of the considered SSM.

2) BIF$_{2}$ is a backward filter associated with a PF (in particular, a 
\emph{sequential importance resampling} filter \cite{Arulampalam_2002})
operating on the nonlinear state component only (so that $\mathbf{x}%
_{l}^{(2)}=\mathbf{x}_{l}^{(N)}$ and $\mathbf{\bar{x}}_{l}^{(2)}=\mathbf{x}%
_{l}^{(L)}$) and representing it through a set of $N_{p}$ particles (note
that $N_{d}=D_{N}$ elements of the system state are shared by the two BIF
algorithms). This means that BIF$_{2}$ is employed to compute new weights
for all the elements of the particle set generated by the PF in the forward
pass.

\begin{figure}[tbp]
\centering
\includegraphics[width=0.70\textwidth]{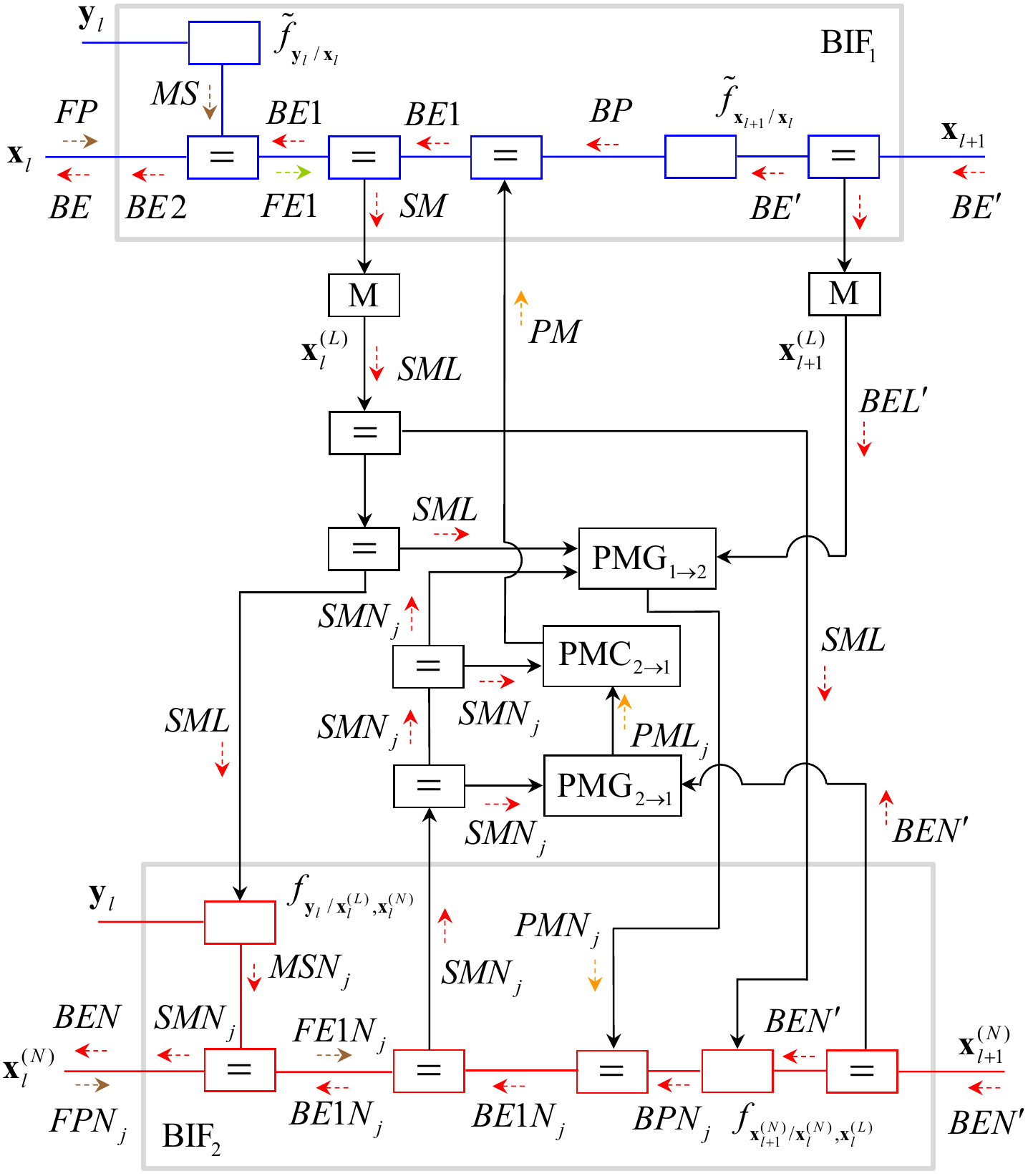}
\caption{Parallel concatenation of two backward information filters, one
paired with an EKF, the other one with a PF.}
\label{Fig_4}
\end{figure}

Based on the general models shown in Figs. \ref{Fig_2} and \ref{Fig_3}, the
specific graphical model illustrated in Fig. \ref{Fig_4} (and referring to
the $(T-l)$-th recursion of backward filtering) can be drawn for the
considered case. In the following, we provide various details about the
adopted notation and the message passing within each constituent filter and
from each filter to the other one.

\emph{Message passing within} BIF$_{1}$ - BIF$_{1}$ is based on the \emph{%
approximate} statistical models $\tilde{f}(\mathbf{x}_{l+1}|\mathbf{x}_{l})$
and $\tilde{f}(\mathbf{y}_{l}|\mathbf{x}_{l})$; these are derived from the
linearised eqs. (\ref{state_up_approx}) and (\ref{meas_mod_approx}),
respectively. Moreover, the (Gaussian) messages passed over its graph
(enclosed within the upper rectangle appearing in Fig. \ref{Fig_4}) are $%
\vec{m}_{fp}(\mathbf{x}_{l})$, $m_{ms}(\mathbf{x}_{l})$, $\vec{m}_{fe1}(%
\mathbf{x}_{l})$, $m_{pm}(\mathbf{x}_{l})$, $\overset{\leftarrow }{m}_{be1}(%
\mathbf{x}_{l})$, $\overset{\leftarrow }{m}_{be2}(\mathbf{x}_{l})$ ($=%
\overset{\leftarrow }{m}_{be}(\mathbf{x}_{l})$), $\overset{\leftarrow }{m}%
_{bp}(\mathbf{x}_{l})$ and $\overset{\leftarrow }{m}_{be}(\mathbf{x}_{l+1})$%
, and are denoted $FP$, $MS$, $FE1$, $PM$, $BE1$, $BE2$ ($BE$), $BP$ and $%
BE^{^{\prime }}$, respectively, to ease reading.

\emph{Message passing within} BIF$_{2}$ - BIF$_{2}$ is based on the \emph{%
exact} statistical models $f(\mathbf{x}_{l+1}^{(N)}|\mathbf{x}_{l}^{(N)}$, $%
\mathbf{x}_{l}^{(L)})$ and $f(\mathbf{y}_{l}|\mathbf{x}_{l}^{(N)},\mathbf{x}%
_{l}^{(L)})$, that are derived from the eqs. (\ref{eq:XL_update}) (with $Z=N$%
) and (\ref{eq:y_t}), respectively. Moreover, the messages processed by it
and appearing in Fig. \ref{Fig_4} refer to the $j$-th particle \emph{%
predicted} in the previous (i.e., in the $(l-1)$-th) recursion of forward
filtering and denoted $\mathbf{x}_{fp,l,j}^{(N)}$, with $j=0,1,...,N_{p}-1$;
such messages are $\vec{m}_{fp,j}(\mathbf{x}_{l}^{(N)})$, $m_{ms,j}(\mathbf{x%
}_{l}^{(N)})$, $\vec{m}_{fe1,j}(\mathbf{x}_{l}^{(N)})$, $m_{pm,j}(\mathbf{x}%
_{l}^{(N)})$, $\overset{\leftarrow }{m}_{be1,j}(\mathbf{x}_{l}^{(N)})$, $%
\overset{\leftarrow }{m}_{be,j}(\mathbf{x}_{l}^{(N)})$, $\overset{\leftarrow 
}{m}_{bp,j}(\mathbf{x}_{l}^{(N)})$ and $\overset{\leftarrow }{m}_{be,j}(%
\mathbf{x}_{l+1}^{(N)})$, and are denoted $FPN_{j}$, $MSN_{j}$, $FE1N_{j}$, $%
PMN_{j}$, $BE1N_{j}$, $BEN_{j}$, $BPN_{j}$ and $BEN_{j}^{^{\prime }}$,
respectively, to ease reading.

\emph{Message passing from} BIF$_{1}$ \emph{to} BIF$_{2}$ - BIF$_{2}$ is fed
by the message $m_{sm}(\mathbf{x}_{l}^{(L)})$ and the message set $%
\{m_{pm,j}(\mathbf{x}_{l}^{(N)})\}$ conveying pseudo-measurement
information; these messages are computed on the basis of the statistical
information made available by BIF$_{1}$. More specifically, on the one hand,
the message $m_{sm}(\mathbf{x}_{l}^{(L)})$ (denoted $SML$) results from the
marginalization of $m_{sm}(\mathbf{x}_{l})$ and is employed for
marginalising the PF state update and measurement models (i.e., $f(\mathbf{x}%
_{l+1}^{(N)}|\mathbf{x}_{l}^{(N)}$, $\mathbf{x}_{l}^{(L)})$ and $f(\mathbf{y}%
_{l}|\mathbf{x}_{l}^{(N)},\mathbf{x}_{l}^{(L)})$, respectively) with respect
to $\mathbf{x}_{l}^{(L)}$. On the other hand, the pseudo-measurement message 
$m_{pm,j}(\mathbf{x}_{l}^{(N)})$ (denoted $PMN_{j}$) is evaluated in the PMG$%
_{1\rightarrow 2}$ block by processing the messages $m_{sm}(\mathbf{x}%
_{l}^{(L)})$ and $\overset{\leftarrow }{m}_{be}(\mathbf{x}_{l+1}^{(L)})$
(denoted $BEL^{\prime }$ and resulting from the marginalization of $\overset{%
\leftarrow }{m}_{be}(\mathbf{x}_{l+1})$), under the assumption that $\mathbf{%
x}_{l}^{(N)}$ is represented by the $j$-th particle (conveyed by the message 
$m_{sm,j}(\mathbf{x}_{l}^{(N)})$).

As illustrated in the Appendix, the computation of the message $m_{pm,j}(%
\mathbf{x}_{l}^{(N)})$ involves the evaluation of the pdf of the random
vector 
\begin{equation}
\mathbf{z}_{l}^{(N)}\triangleq \mathbf{x}_{l+1}^{(L)}-\mathbf{A}%
_{l}^{(L)}\left( \mathbf{x}_{l}^{(N)}\right) \mathbf{x}_{l}^{(L)}\text{,}
\label{z_N_l}
\end{equation}%
defined on the basis of the state update equation (\ref{eq:XL_update}) (with 
$Z=L$) and conditioned on the fact that $\mathbf{x}_{l}^{(N)}=\mathbf{x}%
_{fp,l,j}^{(N)}$. This pdf, which is computed according to the joint
statistical characterization of $\mathbf{x}_{l}^{(L)}$ and $\mathbf{x}%
_{l+1}^{(L)}$ provided by BIF$_{1}$, is conveyed by the message $m_{j}(%
\mathbf{z}_{l}^{(N)})$ (not appearing in Fig. \ref{Fig_4}). Note also that
from eq. (\ref{eq:XL_update}) (with $Z=L$) the equality 
\begin{equation}
\mathbf{z}_{l}^{(N)}=\mathbf{f}_{l}^{(L)}\left( \mathbf{x}_{l}^{(N)}\right) +%
\mathbf{w}_{l}^{(L)}  \label{z_N_l_bis}
\end{equation}%
is easily inferred; the pdf of $\mathbf{z}_{l}^{(N)}$ evaluated on the basis
of the \emph{right-hand side} (RHS) of eq. (\ref{z_N_l_bis}) is denoted $f(%
\mathbf{z}_{l}^{(N)}|\mathbf{x}_{l}^{(N)})$ in the following.

\emph{Message passing from} BIF$_{2}$ \emph{to} BIF$_{1}$ - BIF$_{1}$ is fed
by the message $m_{pm}(\mathbf{x}_{l})$ that, unlike the set $\{m_{pm,j}(%
\mathbf{x}_{l}^{(N)})\}$ passed to BIF$_{2}$, provides pseudo-measurement
information about the \emph{whole state} $\mathbf{x}_{l}$. This message is
generated as follows. The message set $\{m_{sm,j}(\mathbf{x}_{l}^{(N)})\}$
produced by the PF is processed in the PMG$_{2\rightarrow 1}$ block, that
computes the set of $N_{p}$ pseudo-measurement messages $\{m_{pm,j}(\mathbf{x%
}_{l}^{(L)})\}$ referring to the linear state component only. Then, the two
sets $\{m_{pm,j}(\mathbf{x}_{l}^{(L)})\}$ and $\{m_{sm,j}(\mathbf{x}%
_{l}^{(N)})\}$ are merged in the PMC$_{2\rightarrow 1}$ block, where the
information they convey are \emph{converted} into the (single) message $%
m_{pm}(\mathbf{x}_{l})$. Moreover, as illustrated in the Appendix, the
message $m_{pm,j}(\mathbf{x}_{l}^{(L)})$ conveys a sample of the random
vector \cite{Schon_2005}%
\begin{equation}
\mathbf{z}_{l}^{(L)}\triangleq \mathbf{x}_{l+1}^{(N)}-\mathbf{f}%
_{l}^{(N)}\left( \mathbf{x}_{l}^{(N)}\right) ;  \label{eq:z_L_l}
\end{equation}%
such a sample is generated under the assumption that $\mathbf{x}_{l}^{(N)}=%
\mathbf{x}_{fp,l,j}^{(N)}$. The pdf of the random vector $\mathbf{z}%
_{l}^{(L)}$ is evaluated on the basis of the joint statistical
representation of the couple $(\mathbf{x}_{l}^{(N)}$, $\mathbf{x}%
_{l+1}^{(N)})$ produced by BIF$_{2}$ and is conveyed by the message $m_{j}(%
\mathbf{z}_{l}^{(L)})$ (not appearing in Fig. \ref{Fig_4}); note also that
from eq. (\ref{eq:XL_update}) (with $Z=N$) the equality 
\begin{equation}
\mathbf{z}_{l}^{(L)}=\mathbf{A}_{l}^{(N)}\left( \mathbf{x}_{l}^{(N)}\right) 
\mathbf{x}_{l}^{(L)}+\mathbf{w}_{l}^{(N)}  \label{eq:z_L_l_bis}
\end{equation}%
is easily inferred; the pdf of $\mathbf{z}_{l}^{(N)}$ evaluated on the basis
of the RHS of eq. (\ref{eq:z_L_l_bis}) is denoted $f(\mathbf{z}_{l}^{(L)}|%
\mathbf{x}_{l}^{(L)},\mathbf{x}_{l}^{(N)})$ in the following. 

The rationale behind the message passing illustrated above can be summarized
as follows. The message $m_{pm}(\mathbf{x}_{l})$ is extracted from the
statistical information generated by BIF$_{2}$ and is exploited by BIF$_{1}$
to refine its backward estimate of the whole state; moreover, merging this
estimate with the forward estimate $\vec{m}_{fe1}(\mathbf{x}_{l})$ allows to
generate a more accurate statistical representation for $\mathbf{x}_{l}$
and, consequently, for $\mathbf{x}_{l}^{(L)}$ (these are conveyed by $m_{sm}(%
\mathbf{x}_{l})$ and $m_{sm}(\mathbf{x}_{l}^{(L)})$, respectively); finally,
these statistical information are exploited to aid BIF$_{2}$ in the
computation of more refined weights of the particles representing $\mathbf{x}%
_{l}^{(N)}$.

Given the graphical model shown in Fig. \ref{Fig_4} and the messages passed
over it, the derivation of a specific BITF algorithm requires: a) defining
the mathematical structure of the input messages that feed the $(T-l)$-th
recursion of backward filtering and that of the output messages emerging
from both backward filtering and smoothing in the same recursion; b)
describing message scheduling; c) deriving mathematical expressions for all
the computed messages. These issues are analysed in detail in Section \ref%
{sec:Message-Passing}.

\section{Scheduling and Computation of Probabilistic Messages in Turbo
Smoothing Algorithms for CLG Models\label{sec:Message-Passing}}

In this Section, the specific issues raised at the end of the previous
Section and concerning the message passing accomplished over the graphical
model shown in Fig. \ref{Fig_4} are addressed. For this reason, we first
provide various details about a) the messages feeding backward filtering,
and b) the messages emerging from it and from the related smoothing. Then,
we focus on the scheduling of such messages and on their computation. This
allows us to develop two new smoothing techniques, one solving problem 
\textbf{P.1}, the other one problem \textbf{P.2}. Finally, these techniques
are briefly compared with other particle smoothing methods available in the
literature.

\subsection{Input and Output Messages\label{Input_and_output}}

The \emph{input} messages feeding the $(T-l)$-th recursion of backward
filtering are generated in the $l$-th recursion of the paired forward
filtering and in the previous recursion (i.e., in the $(T-l+1)$-th
recursion) of the backward pass. In the following, various details about
such messages are provided.

1. \emph{Input messages evaluated in the forward pass} - A turbo filter,
consisting of an EKF (denoted F$_{1}$) and a PF (denoted F$_{2}$), is
employed in the forward pass of the devised TS algorithms and is run only
once. Therefore, the forward predictions/estimates, provided by F$_{1}$ (F$%
_{2}$) and made available to BIF$_{1}$ (BIF$_{2}$), are expressed by
Gaussian pdfs (sets of weighted particles), each conveyed by a Gaussian
message (by a set of particle-dependent messages). The notation adopted in
the following for these probabilistic information is summarized below.

Filter F$_{1}$ - This filter, in its $(l-1)$-th recursion, computes the 
\emph{forward prediction} of $\mathbf{x}_{l}$, conveyed by the message%
\footnote{%
Considerations similar to the ones expressed for $\vec{m}_{fp}(\mathbf{x}%
_{l})$ (\ref{m_fp_L_MPF}) and $\vec{m}_{fe1}(\mathbf{x}_{l})$ (\ref%
{m_fe_L_MPF}) can be repeated for the messages $\vec{m}_{fp,j}(\mathbf{x}%
_{l}^{(N)})$ (\ref{m_fp_N_MPF}) and $\vec{m}_{fe,j}(\mathbf{x}_{l}^{(N)})$ (%
\ref{m_fe_N_MPF}), respectively, defined below.} (see Fig. \ref{Fig_4}) 
\begin{equation}
\vec{m}_{fp}\left( \mathbf{x}_{l}\right) \triangleq \mathcal{N}\left( 
\mathbf{x}_{l};\mathbf{\eta }_{fp,l},\mathbf{C}_{fp,l}\right) .
\label{m_fp_L_MPF}
\end{equation}%
This message is updated in the $l$-th recursion of F$_{1}$\ on the basis of
the measurement $\mathbf{y}_{l}$. This produces the Gaussian message 
\begin{equation}
\vec{m}_{fe1}\left( \mathbf{x}_{l}\right) \triangleq \mathcal{N}\left( 
\mathbf{x}_{l};\mathbf{\eta }_{fe1,l},\mathbf{C}_{fe1,l}\right) ,
\label{m_fe_L_MPF}
\end{equation}%
representing a \emph{forward estimate} of $\mathbf{x}_{l}$; the covariance
matrix $\mathbf{C}_{fe1,l}$ and the mean vector $\mathbf{\eta }_{fe1,l}$ can
be evaluated on the basis of the associated precision matrix (see \cite[eqs.
(14)-(17)]{Vitetta_2018}) 
\begin{equation}
\mathbf{W}_{fe1,l}=\mathbf{H}_{l}\mathbf{W}_{e}\mathbf{H}_{l}^{T}+\mathbf{W}%
_{fp,l}  \label{W_fe_L}
\end{equation}%
and of the transformed mean vector%
\begin{equation}
\mathbf{w}_{fe1,l}=\mathbf{H}_{l}\mathbf{W}_{e}\left( \mathbf{y}_{l}-\mathbf{%
v}_{l}\right) +\mathbf{w}_{fp,l},  \label{w_fe_L}
\end{equation}%
respectively; here, $\mathbf{W}_{e}\triangleq \mathbf{C}_{e}^{-1}$, $\mathbf{%
W}_{fp,l}\triangleq (\mathbf{C}_{fp,l})^{-1}$ and $\mathbf{w}%
_{fp,l}\triangleq \mathbf{W}_{fp,l}\mathbf{\eta }_{fp,l}$. The message $\vec{%
m}_{fp}(\mathbf{x}_{l})$ (\ref{m_fp_L_MPF}) enters the graphical model
developed for BIF$_{1}$ (see Fig. \ref{Fig_4}) along the half edge referring
to $\mathbf{x}_{l}$.

Filter F$_{2}$ - This filter, in its $(l-1)$-th recursion, computes the
particle set $S_{fp,l}\triangleq \{\mathbf{x}%
_{fp,l,j}^{(N)},j=0,1,...,N_{p}-1\}$, representing a \emph{forward prediction%
} of $\mathbf{x}_{l}^{(N)}$; the weight assigned to the particle $\mathbf{x}%
_{fp,l,j}^{(N)}$ is equal to $1/N_{p}$ for any $j$, since the use of \emph{%
particle resampling} in each recursion is assumed. The statistical
information available about $\mathbf{x}_{fp,l,j}^{(N)}$ are conveyed by the
message%
\begin{equation}
\vec{m}_{fp,j}\left( \mathbf{x}_{l}^{(N)}\right) \triangleq \delta \left( 
\mathbf{x}_{l}^{(N)}-\mathbf{x}_{fp,l,j}^{(N)}\right) ,  \label{m_fp_N_MPF}
\end{equation}%
with $j=0,1,...,N_{p}-1$. The weight of $\mathbf{x}_{fp,l,j}^{(N)}$ (with $%
j=0,1,...,N_{p}-1$) is updated by F$_{2}$ in its $l$-th recursion\ on the
basis of the measurement $\mathbf{y}_{l}$; \ the new weight is denoted $%
w_{fe,l,j}$ and is conveyed by the forward message 
\begin{equation}
\vec{m}_{fe1,j}\left( \mathbf{x}_{l}^{(N)}\right) \triangleq
w_{fe,l,j}\,\delta \left( \mathbf{x}_{l}^{(N)}-\mathbf{x}_{fp,l,j}^{(N)}%
\right) .  \label{m_fe_N_MPF}
\end{equation}%
Note that the message set $\{\vec{m}_{fe1,j}(\mathbf{x}_{l}^{(N)})\}$
represents the \emph{forward estimate} of $\mathbf{x}_{l}^{(N)}$ computed by
F$_{2}$ in its $l$-th recursion and that the message set $\{\vec{m}_{fp,j}(%
\mathbf{x}_{l}^{(N)})\}$ (see eq. (\ref{m_fp_N_MPF})) enters the graphical
model developed for BIF$_{2}$ along the half edge referring to $\mathbf{x}%
_{l}^{(N)}$ (see Fig. \ref{Fig_4}).

2. \emph{Input messages evaluated in the backward pass} - The $(T-l)$-th
recursion of backward filtering is fed by the input messages%
\begin{equation}
\overset{\leftarrow }{m}_{be}\left( \mathbf{x}_{l+1}\right) \triangleq 
\mathcal{N}\left( \mathbf{x}_{l+1};\mathbf{\eta }_{be,l+1},\mathbf{C}%
_{be,l+1}\right)  \label{mess_be_l}
\end{equation}%
and%
\begin{equation}
\overset{\leftarrow }{m}_{be}\left( \mathbf{x}_{l+1}^{(N)}\right) \triangleq
\delta \left( \mathbf{x}_{l+1}^{(N)}-\mathbf{x}_{be,l+1}^{(N)}\right) ,
\label{mess_be_N_l}
\end{equation}%
that convey the pdf of the backward estimate of $\mathbf{x}_{l+1}$ computed
by BIF$_{1}$ and the backward estimate of $\mathbf{x}_{l+1}^{(N)}$ generated
by BIF$_{2}$, respectively, in the previous recursion.

All the input messages described above are processed to compute: 1) the new
backward estimates $\overset{\leftarrow }{m}_{be}(\mathbf{x}_{l})$ and $%
\overset{\leftarrow }{m}_{be}(\mathbf{x}_{l}^{(N)})$, that represent the
outputs emerging from the $(T-l)$-th recursion of backward filtering; 2) the
required smoothed information (in the form of probabilistic messages) by
merging forward and backward messages. In the remaining part of this
Paragraph, some essential information about the structure of such messages
are provided; details about their computation are given in the next
Paragraph.

1. \emph{Computation of backward estimates} - The computation of the message 
$\overset{\leftarrow }{m}_{be}(\mathbf{x}_{l})$ (BIF$_{1}$) and of the
message set $\{\overset{\leftarrow }{m}_{be,j}(\mathbf{x}_{l}^{(N)})\}$ (BIF$%
_{2}$) is accomplished as follows. First, the \emph{backward prediction} 
\begin{equation}
\overset{\leftarrow }{m}_{bp}\left( \mathbf{x}_{l}\right) \triangleq 
\mathcal{N}\left( \mathbf{x}_{l};\mathbf{\eta }_{bp,l},\mathbf{C}%
_{bp,l}\right)  \label{m_bp}
\end{equation}%
of $\mathbf{x}_{l}$ and the message%
\begin{equation}
\overset{\leftarrow }{m}_{bp,j}\left( \mathbf{x}_{l}^{(N)}\right) \triangleq
w_{bp,l,j}  \label{m_bp_N}
\end{equation}%
conveying a \emph{backward weight} for the $j$-th particle $\mathbf{x}%
_{fp,l,j}^{(N)}$ representing $\mathbf{x}_{l}^{(N)}$ (with $%
j=0,1,...,N_{p}-1 $) are computed by BIF$_{1}$ and BIF$_{2}$, respectively.
Then, in BIF$_{1}$, the message $\overset{\leftarrow }{m}_{bp}\left( \mathbf{%
x}_{l}\right) $ (\ref{m_bp}) is merged with the pseudo-measurement message $%
m_{pm}(\mathbf{x}_{l})$ and the measurement message $m_{ms}(\mathbf{x}_{l})$
in order to compute%
\begin{equation}
\overset{\leftarrow }{m}_{be1}\left( \mathbf{x}_{l}\right) \triangleq 
\mathcal{N}\left( \mathbf{x}_{l};\mathbf{\eta }_{be1,l},\mathbf{C}%
_{be1,l}\right)  \label{m_be1}
\end{equation}%
and (see eq. (\ref{mess_be_l}))%
\begin{equation}
\overset{\leftarrow }{m}_{be2}\left( \mathbf{x}_{l}\right) \triangleq 
\mathcal{N}\left( \mathbf{x}_{l};\mathbf{\eta }_{be2,l},\mathbf{C}%
_{be2,l}\right) =\overset{\leftarrow }{m}_{be}\left( \mathbf{x}_{l}\right) ,
\label{m_be2}
\end{equation}%
respectively. Similarly, in BIF$_{2}$, the message $\overset{\leftarrow }{m}%
_{bp,j}(\mathbf{x}_{l}^{(N)})$ (\ref{m_bp_N}) is merged first with the
pseudo-measurement message $m_{pm,j}(\mathbf{x}_{l}^{(N)})$ in order to
produce the message (see eq. (\ref{mess_be_N_l}))%
\begin{equation}
\overset{\leftarrow }{m}_{be1,j}\left( \mathbf{x}_{l}^{(N)}\right)
\triangleq w_{be1,l,j}  \label{m_be1_N}
\end{equation}%
conveying a new weight for the $j$-th particle $\mathbf{x}_{fp,l,j}^{(N)}$.
Then, the information conveyed by the message set $\{\overset{\leftarrow }{m}%
_{be1,j}(\mathbf{x}_{l}^{(N)})\}$ is merged with that provided by the
measurement-based set $\{m_{ms,j}(\mathbf{x}_{l}^{(N)})\}$ in order to
evaluate the message (see eq. (\ref{mess_be_N_l})) 
\begin{equation}
\overset{\leftarrow }{m}_{be}\left( \mathbf{x}_{l}^{(N)}\right) =\delta
\left( \mathbf{x}_{l}^{(N)}-\mathbf{x}_{be,l}^{(N)}\right) \text{,}
\label{m_be2_N}
\end{equation}%
that conveys a (particle-independent) backward estimate of $\mathbf{x}%
_{l}^{(N)}$.

2. \emph{Computation of smoothed information} - In our work, the evaluation
of smoothed information is based on the same conceptual approach as \cite%
{Vitetta_2018}, \cite{Fong_2002} and \cite{Lindsten_2016}. In fact, the
proposed method is based on the following ideas:

a) The \emph{joint} smoothing pdf $f(\mathbf{x}_{1:T}|\mathbf{y}_{1:T})$ is
estimated by providing multiple (say, $M$) \emph{realizations} of it and a
single realization (i.e., a single \emph{smoothed} state trajectory) is
computed in each backward pass; consequently, generating the smoothing
output requires running a single forward pass and $M$ distinct backward
passes.

b) The factorisation (\ref{factorisation3b}) is exploited to evaluate
smoothed information, i.e. to merge the statistical information emerging
from the forward pass with that computed in any of the $M$ backward passes.
In particular, this formula is employed to combine the statistical
information made available by F$_{1}$ (F$_{2}$) with those generated by BIF$%
_{1}$ (BIF$_{2}$); consequently, the first factor and the second one
appearing in the RHS of\ eq. (\ref{factorisation3b}) are expressed by the
forward message $\vec{m}_{fe1}(\mathbf{x}_{l})$ (\ref{m_fe_L_MPF}) and the
backward message $\overset{\leftarrow }{m}_{bp}\left( \mathbf{x}_{l}\right) $
(\ref{m_bp}) (the forward message $\vec{m}_{fe1,j}(\mathbf{x}_{l}^{(N)})$ (%
\ref{m_fe_N_MPF}) and the backward message $\overset{\leftarrow }{m}_{bp,j}(%
\mathbf{x}_{l}^{(N)})$ (\ref{m_bp_N})), respectively, if F$_{1}$ and BIF$%
_{1} $ (F$_{2}$ and BIF$_{2}$) are considered.

\subsection{Scheduling and Computation of Probabilistic Messages\label%
{message_passing_algorithms}}

\begin{figure}[tbp]
\centering
\includegraphics[width=0.70\textwidth]{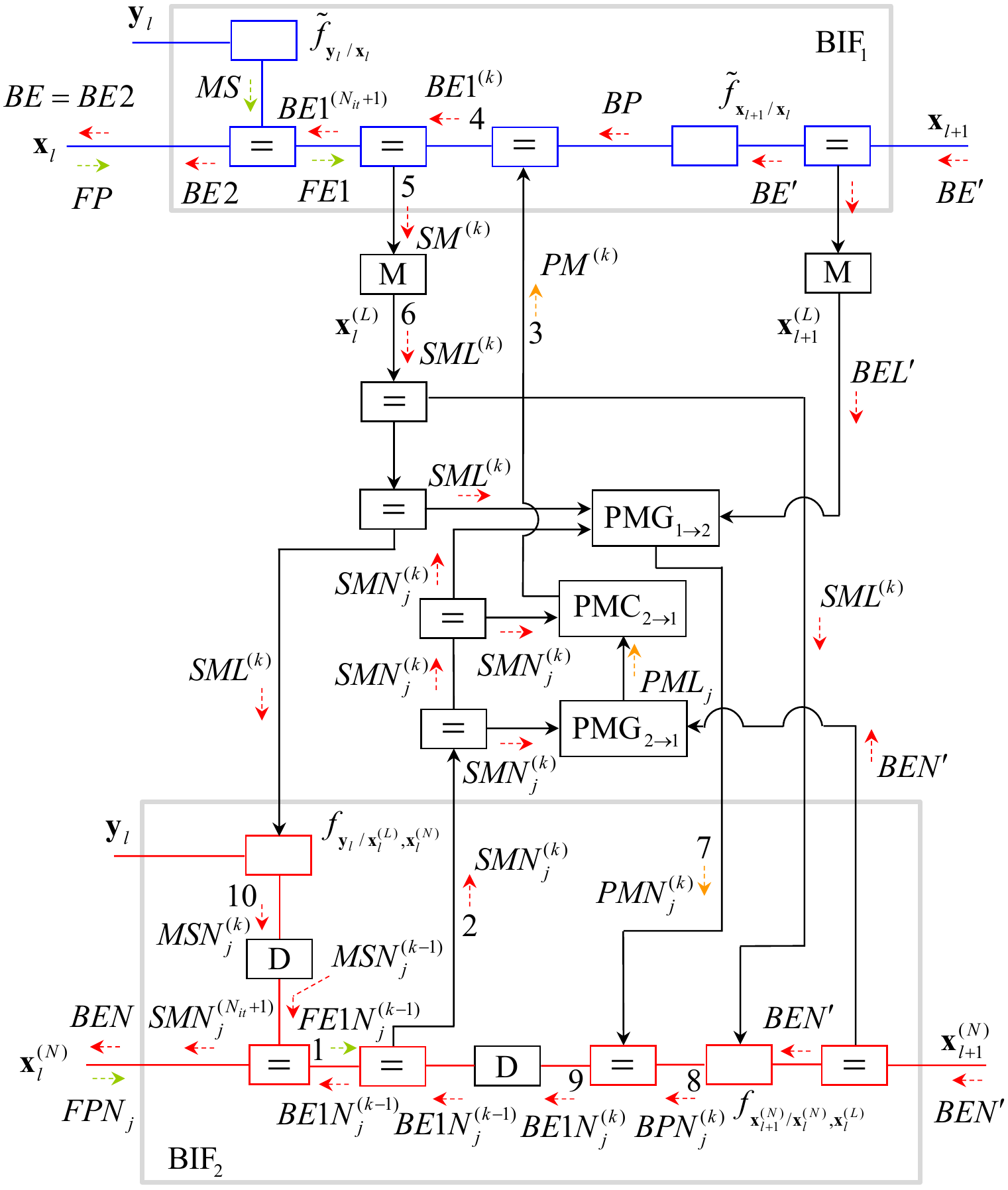}
\caption{Representation of the message scheduling employed in the $k$-th
iteration accomplished within the $(T-l)$-th recursion of BITF\ and TS; the
integers $1-10$ specify the order according to which messages are computed
in the considered iteration. Brown and red arrows are employed to identify
the input/output backward messages and the remaining messages, respectively.}
\label{Fig_5}
\end{figure}

The message passing algorithm we propose for backward filtering and
smoothing is iterative, since, within each recursion of the backward pass,
it can accomplish multiple passes over the same edges. Moreover, it results
from: a) the adoption of the \emph{message scheduling} illustrated in Fig. %
\ref{Fig_5}, that refers to the $k$-th iteration of the devised algorithm;
b) the use of the SPA\ in the evaluation of all the passed messages. It is
also important to mention that the selected scheduling mimics the one
employed in \cite{Vitetta_2018}, which, in turn, has been inspired by \cite%
{Fong_2002} and \cite{Lindsten_2016}. Based on this scheduling, the
computation of the messages passed over the given graphical model can be
divided in the three consecutive phases listed below.

\textbf{I} - In this phase, $\overset{\leftarrow }{m}_{be}(\mathbf{x}_{l+1})$
($BE^{\prime }$) is processed to compute $\overset{\leftarrow }{m}_{bp}(%
\mathbf{x}_{l})$ ($BP$) and $\overset{\leftarrow }{m}_{be}(\mathbf{x}%
_{l+1}^{(L)})$ ($BEL^{\prime }$); moreover, the set $\{m_{pm,j}(\mathbf{x}%
_{l}^{(L)})\}$ ($PML_{j}$) conveying pseudo-measurement information about $%
\mathbf{x}_{l}^{(L)}$ is evaluated.

\textbf{II} - In the second phase, an iterative evaluation of the backward
estimates of the whole state (BIF$_{1}$) and of the nonlinear state
component (BIF$_{2}$) is accomplished. More specifically, in the $k$-th
iteration of this procedure (with $k=1,2,...,N_{it}$, where $N_{it}$ is the
overall number of iterations) the ordered computation of the following
messages or sets of $N_{p}$ messages is accomplished in five consecutive
steps\footnote{%
Note that the superscript $(k)$ ($(k-1)$) indicates that the associated
message is computed in the $k$-th ($(k-1)$-th) iteration of phase II.} (see
Fig. \ref{Fig_5}): 1) $\{m_{sm,j}^{(k)}(\mathbf{x}_{l}^{(N)})\}$ ($%
SMN_{j}^{(k)}$), $m_{pm}^{(k)}(\mathbf{x}_{l})$ ($PM^{(k)}$); 2) $\overset{%
\leftarrow }{m}_{be1}^{(k)}(\mathbf{x}_{l})$ ($BE1^{(k)}$), $m_{sm}^{(k)}(%
\mathbf{x}_{l})$ ($SM^{(k)}$), $m_{sm}^{(k)}(\mathbf{x}_{l}^{(L)})$ ($%
SML^{(k)}$); 3) $\{m_{pm,j}^{(k)}(\mathbf{x}_{l}^{(N)})\}$ ($PMN_{j}^{(k)}$%
); 4) $\{\overset{\leftarrow }{m}_{bp,j}^{(k)}(\mathbf{x}_{l}^{(N)})\}$ ($%
BPN_{j}^{(k)}$), $\{\overset{\leftarrow }{m}_{be1,j}^{(k)}(\mathbf{x}%
_{l}^{(N)})\}$ ($BE1N_{j}^{(k)}$); 5) $\{m_{ms,j}^{(k)}(\mathbf{x}%
_{l}^{(N)})\}$ ($MSN_{j}^{(k)}$).

\textbf{III} - In the third phase, the smoothed information $%
\{m_{sm,j}^{(N_{it}+1)}(\mathbf{x}_{l}^{(N)})\}$ is computed and employed in
the evaluation of: a) the output message $m_{be}(\mathbf{x}_{l}^{(N)})$ of
BIF$_{1}$; b) the new pseudo-measurement message $m_{pm}^{(N_{it}+1)}(%
\mathbf{x}_{l})$. Finally, $m_{pm}^{(N_{it}+1)}(\mathbf{x}_{l})$ is
processed to compute $\overset{\leftarrow }{m}_{be1,l}^{(N_{it}+1)}\left( 
\mathbf{x}_{l}\right) $ and the output message $\overset{\leftarrow }{m}%
_{be,l}\left( \mathbf{x}_{l}\right) =$ $\overset{\leftarrow }{m}%
_{be2,l}\left( \mathbf{x}_{l}\right) $ of BIF$_{2}$.

In the remaining part of this Section, the expressions of all the messages
computed in each of the three phases described above are provided; the
derivation of these expressions is sketched in the Appendix.

\textbf{Phase I }- The message $\overset{\leftarrow }{m}_{be}(\mathbf{x}%
_{l+1}^{(L)})$ is computed as 
\begin{eqnarray}
\overset{\leftarrow }{m}_{be}\left( \mathbf{x}_{l+1}^{(L)}\right)
&\triangleq &\int \overset{\leftarrow }{m}_{be}(\mathbf{x}_{l+1})\,d\mathbf{x%
}_{l+1}^{(N)}  \notag \\
&=&\mathcal{N(}\mathbf{x}_{l+1}^{(L)};\mathbf{\tilde{\eta}}_{be,l+1},\mathbf{%
\tilde{C}}_{be,l+1}),  \label{m_be_x_L_l+1}
\end{eqnarray}%
since it results from the marginalization of $\overset{\leftarrow }{m}_{be}(%
\mathbf{x}_{l+1})$ (\ref{mess_be_l}) with respect to $\mathbf{x}_{l+1}^{(N)}$%
; in practice, the mean vector $\mathbf{\tilde{\eta}}_{be,l+1}$ and the
covariance matrix $\mathbf{\tilde{C}}_{be,l+1}$ are extracted from the
parameters $\mathbf{\eta }_{be,l+1}$ and $\mathbf{C}_{be,l+1}$, respectively
(since $\mathbf{x}_{l+1}^{(L)}$ consists of the first $D_{L}$ elements of $%
\mathbf{x}_{l+1}$).

The message $\overset{\leftarrow }{m}_{bp}(\mathbf{x}_{l})$ (\ref{m_bp}),
representing a \emph{one-step backward prediction} of $\mathbf{x}_{l}$, is
computed on the basis of $\overset{\leftarrow }{m}_{be}(\mathbf{x}_{l+1})$
and the pdf $f(\mathbf{x}_{l+1}|\mathbf{x}_{l})$. Its parameters $\mathbf{%
\eta }_{bp,l}$ and $\mathbf{C}_{bp,l}$ are evaluated on the basis of the
precision matrix%
\begin{equation}
\mathbf{W}_{bp,l}\triangleq \left( \mathbf{C}_{bp,l}\right) ^{-1}=\mathbf{F}%
_{l}^{T}\mathbf{P}_{l+1}\mathbf{W}_{be,l+1}\mathbf{F}_{l}  \label{W_bp_x_l}
\end{equation}%
and of the transformed mean vector%
\begin{equation}
\mathbf{w}_{bp,l}\triangleq \mathbf{W}_{bp,l}\mathbf{\eta }_{bp,l}=\mathbf{F}%
_{l}^{T}[\mathbf{P}_{l+1}\mathbf{w}_{be,l+1}-\mathbf{W}_{be,l+1}\mathbf{Q}%
_{l+1}\mathbf{W}_{w}\mathbf{u}_{l}],  \label{w_bp_x_l}
\end{equation}%
respectively; here, $\mathbf{W}_{be,l+1}\triangleq (\mathbf{C}%
_{be,l+1})^{-1} $, $\mathbf{P}_{l+1}\triangleq \mathbf{\mathbf{I}}_{D}-%
\mathbf{W}_{be,l+1}\mathbf{Q}_{l+1}$, $\mathbf{Q}_{l+1}\triangleq (\mathbf{W}%
_{w}+\mathbf{W}_{be,l+1})^{-1}$, $\mathbf{W}_{w}\triangleq (\mathbf{C}%
_{w})^{-1}$ and $\mathbf{w}_{be,l+1}\triangleq \mathbf{W}_{be,l+1}\mathbf{%
\eta }_{be,l+1}$.

The evaluation of the set of messages $\{m_{pm,j}(\mathbf{x}_{l}^{(L)})\}$
is based on the message $\overset{\leftarrow }{m}_{be}(\mathbf{x}%
_{l+1}^{(N)})$ (\ref{mess_be_N_l}) and on the particle set conveyed by the
messages $\{m_{sm,j}^{(k)}(\mathbf{x}_{l}^{(N)})\}$ (such a set, being equal
to $S_{fp,l}$, is independent of the iteration index $k$; see eq. (\ref%
{m_sm_j_N2})). In the Appendix it is shown that%
\begin{equation}
m_{pm,j}\left( \mathbf{x}_{l}^{(L)}\right) =\mathcal{\mathcal{N}}\left( 
\mathbf{x}_{l}^{(L)};\mathbf{\tilde{\eta}}_{pm,l,j},\mathbf{\tilde{C}}%
_{pm,l,j}\right) ;  \label{eq:message_pm_L_j_tris}
\end{equation}%
the covariance matrix $\mathbf{\tilde{C}}_{pm,l,j}$ and the mean vector $%
\mathbf{\tilde{\eta}}_{pm,l,j}$ are computed on the basis of the precision
matrix%
\begin{equation}
\mathbf{\tilde{W}}_{pm,l,j}\triangleq \left( \mathbf{\tilde{C}}%
_{pm,l,j}\right) ^{-1}=\left( \mathbf{A}_{l,j}^{(N)}\right) ^{T}\mathbf{W}%
_{w}^{(N)}\mathbf{A}_{l,j}^{(N)}  \label{eq:W_pm_L_j}
\end{equation}%
and of the transformed mean vector%
\begin{equation}
\mathbf{\tilde{w}}_{pm,l,j}\triangleq \mathbf{\tilde{W}}_{pm,l,j}\mathbf{%
\tilde{\eta}}_{pm,l,j}=\left( \mathbf{A}_{l,j}^{(N)}\right) ^{T}\mathbf{W}%
_{w}^{(N)}\mathbf{z}_{l,j}^{(L)},  \label{eq:w_pm_L_j}
\end{equation}%
respectively; here, $\mathbf{A}_{l,j}^{(N)}\triangleq \mathbf{A}_{l}^{(N)}(%
\mathbf{x}_{fp,l,j}^{(N)})$, 
\begin{equation}
\mathbf{z}_{l,j}^{(L)}\triangleq \mathbf{x}_{be,l+1}^{(N)}-\mathbf{f}%
_{l,j}^{(N)}  \label{PM_z_L}
\end{equation}%
is an iteration-independent pseudo-measurement and $\mathbf{f}%
_{l,j}^{(N)}\triangleq \mathbf{f}_{l}^{(N)}(\mathbf{x}_{fp,l,j}^{(N)})$.

\textbf{Phase II} - A short description of the five steps accomplished in
the $k$-th iteration of this phase is provided in the following.

Step 1) \emph{Computation of the pseudo-measurements for} BIF$_{1}$- The
message $m_{sm,j}^{(k)}(\mathbf{x}_{l}^{(N)})$ is evaluated as\footnote{%
Note that the messages $\overset{\rightarrow }{m}_{fe1,j}^{(k-1)}(\mathbf{x}%
_{l}^{(N)})\,$\ and $\overset{\leftarrow }{m}_{be1,j}^{(k-1)}(\mathbf{x}%
_{l}^{(N)})$ appearing in the following formula are evaluated in the
previous iteration and stored in the \emph{delay} elements (identified by
the letter D in Fig. \ref{Fig_5}).} (see Fig. \ref{Fig_5}, and eqs. (\ref%
{m_fe_N_MPF}) and (\ref{m_be1_N}))%
\begin{eqnarray}
m_{sm,j}^{(k)}\left( \mathbf{x}_{l}^{(N)}\right) &=&\overset{\rightarrow }{m}%
_{fe1,j}^{(k-1)}\left( \mathbf{x}_{l}^{(N)}\right) \,\overset{\leftarrow }{m}%
_{be1,j}^{(k-1)}\left( \mathbf{x}_{l}^{(N)}\right)  \label{m_sm_j_N1} \\
&=&w_{sm,l,j}^{(k)}\,\delta \left( \mathbf{x}_{l}^{(N)}-\mathbf{x}%
_{fp,l,j}^{(N)}\right) ,  \label{m_sm_j_N2}
\end{eqnarray}%
where 
\begin{equation}
~w_{sm,l,j}^{(k)}\triangleq w_{fe1,l,j}^{(k-1)}\,w_{be1,l,j}^{(k-1)},
\label{w__sm_j_N}
\end{equation}%
with $w_{fe1,l,j}^{(0)}=w_{fe,l,j}$ (see eq. (\ref{m_fe_N_MPF})) and $%
w_{be1,l,j}^{(0)}=1$ (i.e., $w_{sm,l,j}^{(1)}=w_{fe,l,j}$). Then, the
weights $\{w_{sm,l,j}^{(k)}\}$ are normalized; this produces the $j$-th
normalised weight%
\begin{equation}
W_{sm,l,j}^{(k)}\triangleq K_{sm,l}^{(k)}\,w_{sm,l,j}^{(k)}\,,
\label{W_fe_2_x_N_l}
\end{equation}%
with $j=0,1,...,N_{p}-1$, where $K_{sm,l}^{(k)}\triangleq
1/\sum\limits_{j=0}^{N_{p}-1}w_{sm,l,j}^{(k)}$. Note that the particles $\{%
\mathbf{x}_{fp,l,j}^{(N)}\}$ and their new weights $\{W_{sm,l,j}^{(k)}\}$
provide a statistical representation of the \emph{smoothed estimate} of $%
\mathbf{x}_{l}^{(N)}$ evaluated in the $k$-th iteration.

Then, the message 
\begin{equation}
m_{pm}^{(k)}(\mathbf{x}_{l})=\mathcal{\mathcal{N}}\left( \mathbf{x}_{l};%
\mathbf{\eta }_{pm,l}^{(k)},\mathbf{C}_{pm,l}^{(k)}\right)  \label{m_pm_x_l}
\end{equation}%
is computed in the block PMC$_{2\rightarrow 1}$ on the basis of the message
sets $\{m_{pm,j}(\mathbf{x}_{l}^{(L)})\}$ (see eq. (\ref%
{eq:message_pm_L_j_tris})) and $\{m_{sm,j}^{(k)}(\mathbf{x}_{l}^{(N)})\}$;
the mean vector $\mathbf{\eta }_{pm,l}^{(k)}$ and the covariance matrix $%
\mathbf{C}_{pm,l}^{(k)}$ are evaluated as 
\begin{equation}
\mathbf{\eta }_{pm,l}^{(k)}=\left[ \left( \mathbf{\eta }_{pm,l}^{(L,k)}%
\right) ^{T},\left( \mathbf{\eta }_{pm,l}^{(N,k)}\right) ^{T}\right] ^{T}
\label{eta_pm_l_k}
\end{equation}%
and%
\begin{equation}
\mathbf{C}_{pm,l}^{(k)}=\left[ 
\begin{array}{cc}
\mathbf{C}_{pm,l}^{(LL,k)} & \mathbf{C}_{pm,l}^{(LN,k)} \\ 
\left( \mathbf{C}_{pm,l}^{(LN,k)}\right) ^{T} & \mathbf{C}_{pm,l}^{(NN,k)}%
\end{array}%
\right] ,  \label{C_pm_l_k}
\end{equation}%
respectively, where%
\begin{equation}
\mathbf{\eta }_{pm,l}^{(X,k)}\triangleq
\sum_{j=0}^{N_{p}-1}W_{sm,l,j}^{(k)}\,\mathbf{\eta }_{pm,l,j}^{(X)}
\label{eta_pm_l_L_k}
\end{equation}%
is a $D_{X}$-dimensional mean vector (with $X=L$ and $N)$, 
\begin{equation}
\mathbf{C}_{pm,l}^{(XY,k)}\triangleq \sum_{j=0}^{N_{p}-1}W_{sm,l,j}^{(k)}%
\mathbf{r}_{pm,l,j}^{(XY)}-\mathbf{\eta }_{pm,l}^{(X)}\left( \mathbf{\eta }%
_{pm,l}^{(Y)}\right) ^{T}  \label{C_pm_l_L_k_bis}
\end{equation}%
is a $D_{X}\times D_{Y}$ covariance (or cross-covariance) matrix (with $%
XY=LL $, $NN$ and $LN)$, $\mathbf{\eta }_{pm,l,j}^{(L)}=\mathbf{\tilde{\eta}}%
_{pm,l,j}$, $\mathbf{\eta }_{pm,l,j}^{(N)}=\mathbf{x}_{fp,l,j}^{(N)}$, $%
\mathbf{r}_{pm,l,j}^{(LL)}\triangleq \mathbf{\tilde{C}}_{pm,l,j}+\mathbf{%
\tilde{\eta}}_{pm,l,j}(\mathbf{\tilde{\eta}}_{pm,l,j})^{T}$, $\mathbf{r}%
_{pm,l,j}^{(NN)}\triangleq \mathbf{x}_{fp,l,j}^{(N)}(\mathbf{x}%
_{fp,l,j}^{(N)})^{T}$ and $\mathbf{r}_{pm,l,j}^{(LN)}\triangleq \mathbf{%
\tilde{\eta}}_{pm,l,j}(\mathbf{x}_{fp,l,j}^{(N)})^{T}$.

Step 2) \emph{Computation of the backward and smoothed estimates in} BIF$%
_{1} $ - The message $\overset{\leftarrow }{m}_{be1}^{(k)}(\mathbf{x}_{l})$
is evaluated as (see Fig. \ref{Fig_5}) 
\begin{eqnarray}
\overset{\leftarrow }{m}_{be1}^{(k)}(\mathbf{x}_{l}) &=&\overset{\leftarrow }%
{m}_{bp}(\mathbf{x}_{l})\,m_{pm}^{(k)}(\mathbf{x}_{l})  \label{m_be1_x_la} \\
&=&\mathcal{\mathcal{N}}\left( \mathbf{x}_{l};\mathbf{\eta }_{be1,l}^{(k)},%
\mathbf{C}_{be1,l}^{(k)}\right) ,  \label{m_be1_x_laa}
\end{eqnarray}%
where the messages $\overset{\leftarrow }{m}_{bp}(\mathbf{x}_{l})$ and $%
m_{pm}^{(k)}(\mathbf{x}_{l})$ are given by eq. (\ref{m_bp}) and eq. (\ref%
{m_pm_x_l}), respectively. The covariance matrix $\mathbf{C}_{be1,l}^{(k)}$
and the mean vector $\mathbf{\eta }_{be1,l}^{(k)}$ are computed on the basis
of the associated precision matrix 
\begin{equation}
\mathbf{W}_{be1,l}^{(k)}\triangleq (\mathbf{C}_{be1,l}^{(k)})^{-1}=\mathbf{W}%
_{bp,l}+\mathbf{W}_{pm,l}^{(k)}  \label{W_be1_l_kn}
\end{equation}%
and transformed mean vector%
\begin{equation}
\mathbf{w}_{be1,l}^{(k)}\triangleq \mathbf{W}_{be1,l}^{(k)}\mathbf{\eta }%
_{be1,l}^{(k)}=\mathbf{w}_{bp,l}+\mathbf{w}_{pm,l}^{(k)},  \label{w_be1_l_kn}
\end{equation}%
respectively; here, $\mathbf{W}_{pm,l}^{(k)}\triangleq (\mathbf{C}%
_{pm,l}^{(k)})^{-1}$, $\mathbf{w}_{pm,l}^{(k)}\triangleq \mathbf{W}%
_{pm,l}^{(k)}\,\mathbf{\eta }_{pm,l}^{(k)}$, and $\mathbf{W}_{bp,l}$ and $%
\mathbf{w}_{bp,l}$ are given by eqs. (\ref{W_bp_x_l}) and (\ref{w_bp_x_l}),
respectively. From eqs. (\ref{W_be1_l_kn})-(\ref{w_be1_l_kn}) the
expressions 
\begin{equation}
\mathbf{C}_{be1,l}^{(k)}=\mathbf{W}_{l}^{(k)}\mathbf{C}_{pm,l}^{(k)}
\label{C_be1_l_ka}
\end{equation}%
and%
\begin{equation}
\mathbf{\eta }_{be1,l}^{(k)}=\mathbf{W}_{l}^{(k)}\left[ \mathbf{C}%
_{pm,l}^{(k)}\mathbf{w}_{bp,l}+\mathbf{\eta }_{pm,l}^{(k)}\right]
\label{eta_be1_l_ka}
\end{equation}%
can be easily inferred; here, $\mathbf{W}_{l}^{(k)}\triangleq \lbrack 
\mathbf{C}_{pm,l}^{(k)}\mathbf{W}_{bp,l}+\mathbf{I}_{D}]^{-1}$.

Then, the message $m_{sm}^{(k)}(\mathbf{x}_{l})$ is evaluated as (see Fig. %
\ref{Fig_5}) 
\begin{eqnarray}
m_{sm}^{(k)}\left( \mathbf{x}_{l}\right) &=&\vec{m}_{fe1}\left( \mathbf{x}%
_{l}\right) \overset{\leftarrow }{m}_{be1}^{(k)}\left( \mathbf{x}_{l}\right)
\label{m_sm_x_la} \\
&=&\mathcal{\mathcal{N}}\left( \mathbf{x}_{l};\mathbf{\eta }_{sm,l}^{(k)},%
\mathbf{C}_{sm,l}^{(k)}\right) ,  \label{m_sm_x_l}
\end{eqnarray}%
where the messages $\vec{m}_{fe1}\left( \mathbf{x}_{l}\right) $ and $\overset%
{\leftarrow }{m}_{be1}^{(k)}\left( \mathbf{x}_{l}\right) $ are given by eqs.
(\ref{m_fe_L_MPF}) and (\ref{m_be1_x_laa}), respectively. The covariance
matrix $\mathbf{C}_{sm,l}^{(k)}$ and the mean vector $\mathbf{\eta }%
_{be1,l}^{(k)}$ are computed on the basis of the associated precision matrix%
\begin{equation}
\mathbf{W}_{sm,l}^{(k)}=\mathbf{W}_{fe1,l}+\mathbf{W}_{be1,l}^{(k)}
\label{W_sm_l_k}
\end{equation}%
and transformed mean vector%
\begin{equation}
\mathbf{w}_{sm,l}^{(k)}=\mathbf{w}_{fe1,l}+\mathbf{w}_{be1,l}^{(k)},
\label{w_sm_l_k}
\end{equation}%
respectively. Finally, marginalizing $m_{sm}^{(k)}(\mathbf{x}_{l})$ (\ref%
{m_sm_x_l}) with respect to $\mathbf{x}_{l}^{(N)}$ results in the message 
\begin{equation}
m_{sm}^{(k)}\left( \mathbf{x}_{l}^{(L)}\right) \triangleq \int m_{sm}^{(k)}(%
\mathbf{x}_{l})d\mathbf{x}_{l}^{(N)}=\mathcal{N(}\mathbf{x}_{l}^{(L)};%
\mathbf{\tilde{\eta}}_{sm,l}^{(k)},\mathbf{\tilde{C}}_{sm,l}^{(k)}),
\label{m_fe_L_EKF_2}
\end{equation}%
where $\mathbf{\tilde{\eta}}_{sm,l}^{(k)}$ and $\mathbf{\tilde{C}}%
_{sm,l}^{(k)}$ are extracted from the mean $\mathbf{\eta }_{sm,l}^{(k)}$ and
the covariance matrix $\mathbf{C}_{sm,l}^{(k)}$ of $m_{sm}^{(k)}(\mathbf{x}%
_{l})$ (\ref{m_sm_x_l}), respectively (since $\mathbf{x}_{l}^{(L)}$ consists
of the first $D_{L}$ elements of $\mathbf{x}_{l}$).

Step 3) \emph{Computation of the pseudo-measurements for} BIF$_{2}$\textbf{\ 
}- The pseudo-measurement information feeding BIF$_{2}$ is conveyed by the
message set $\{m_{pm,j}^{(k)}(\mathbf{x}_{l}^{(N)})\triangleq
w_{pm,l,j}^{(k)}\}$, i.e. by a set of new weights for the particles forming
the set $S_{fp,l}$. The $j$-th weight is evaluated as 
\begin{equation}
w_{pm,l,j}^{(k)}=D_{pm,l,j}^{(k)}\exp \left( -\frac{1}{2}Z_{pm,l,j}^{(k)}%
\right)  \label{m_pm_x_N_l_j}
\end{equation}%
for any $j$; here, 
\begin{equation}
Z_{pm,l,j}^{(k)}\triangleq \left\Vert \mathbf{\check{\eta}}%
_{z,l,j}^{(k)}\right\Vert _{\mathbf{\check{W}}_{z,l,j}^{(k)}}^{2}+\left\Vert 
\mathbf{f}_{l,j}^{(L)}\right\Vert _{\mathbf{W}_{w}^{(L)}}^{2}-\left\Vert 
\mathbf{\check{\eta}}_{pm,l,j}^{(k)}\right\Vert _{\mathbf{\check{W}}%
_{pm,l,j}^{(k)}}^{2},  \label{Z_pm}
\end{equation}%
$\left\Vert \mathbf{x}\right\Vert _{\mathbf{W}}^{2}\triangleq \mathbf{x}^{T}%
\mathbf{Wx}$ denotes the square of the norm of the vector $\mathbf{x}$ with
respect to the positive definite matrix $\mathbf{W}$,%
\begin{equation}
\mathbf{\check{W}}_{pm,l,j}^{(k)}\triangleq \left( \mathbf{\check{C}}%
_{pm,l,j}^{(k)}\right) ^{-1}=\mathbf{\check{W}}_{z,l,j}^{(k)}+\mathbf{W}%
_{w}^{(L)},  \label{W_pm_x_N_l_j}
\end{equation}%
\begin{equation}
\mathbf{\check{w}}_{pm,l,j}^{(k)}\triangleq \mathbf{\check{W}}_{pm,l,j}^{(k)}%
\mathbf{\check{\eta}}_{pm,l,j}^{(k)}=\mathbf{\check{w}}_{z,l,j}^{(k)}+%
\mathbf{W}_{w}^{(L)}\mathbf{f}_{l,j}^{(L)},  \label{w_pm_x_N_l_j}
\end{equation}%
$\mathbf{\check{W}}_{z,l,j}^{(k)}\triangleq (\mathbf{\check{C}}%
_{z,l,j}^{(k)})^{-1}$, $\mathbf{\check{w}}_{z,l,j}^{(k)}\triangleq \mathbf{%
\check{W}}_{z,l,j}^{(k)}\mathbf{\check{\eta}}_{z,l,j}^{(k)}$, $\mathbf{%
\check{\eta}}_{z,l,j}^{(k)}$ and $\mathbf{\check{C}}_{z,l,j}^{(k)}$ are
expressed by eqs. (\ref{eta_mess_z_N}) and (\ref{C_mess_Z_N_bis}),
respectively, $\mathbf{W}_{w}^{(L)}\triangleq \lbrack \mathbf{C}%
_{w}^{(L)}]^{-1}$, $\mathbf{f}_{l,j}^{(L)}\triangleq \mathbf{f}_{l}^{(L)}(%
\mathbf{x}_{fp,l,j}^{(N)})$, $D_{pm,l,j}^{(k)}\triangleq \lbrack \det (%
\mathbf{\check{C}}_{l,j}^{(k)})]^{-D_{L}/2}$ and $\mathbf{\check{C}}%
_{l,j}^{(k)}\triangleq \mathbf{\check{C}}_{z,l,j}^{(k)}+\mathbf{C}_{w}^{(L)}$%
.

Step 4) \emph{Computation of the backward weights in} BIF$_{2}$ -\ The
backward message $\overset{\leftarrow }{m}_{bp,j}^{(k)}(\mathbf{x}%
_{l}^{(N)}) $ (\ref{m_bp_N}), i.e. the backward weight (see Fig. \ref{Fig_5}%
) is computed as%
\begin{eqnarray}
w_{bp,l,j}^{(k)} &=&\int \int f(\mathbf{x}_{l+1}^{(N)}/\mathbf{x}%
_{fp,l,j}^{(N)},\mathbf{x}_{l}^{(L)})\overset{\leftarrow }{m}_{be}\left( 
\mathbf{x}_{l+1}^{(N)}\right)  \notag \\
&&\cdot m_{sm}^{(k)}\left( \mathbf{x}_{l}^{(L)}\right) d\mathbf{x}_{l}^{(N)}d%
\mathbf{x}_{l}^{(N)}  \label{weight_bp} \\
&=&D_{bp,l,j}^{(k)}\exp \left( -\frac{1}{2}Z_{bp,l,j}^{(k)}\right) =\overset{%
\leftarrow }{m}_{bp,j}^{(k)}(\mathbf{x}_{l}^{(N)}),  \label{weight_bpa}
\end{eqnarray}%
where $D_{bp,l,j}^{(k)}=(2\pi \det (\mathbf{C}_{1,l,j}^{(N)}))^{-D_{N}/2}$,%
\begin{equation}
Z_{bp,l,j}^{(k)}\triangleq \left\Vert \mathbf{x}_{be,l+1}^{(N)}-\mathbf{\eta 
}_{1,l,j}^{(N)}[k]\right\Vert _{\mathbf{W}_{1,l,j}^{(N)}[k]}^{2}
\label{Z_bp}
\end{equation}%
\begin{equation}
\mathbf{\eta }_{1,l,j}^{(N)}[k]\triangleq \mathbf{A}_{l,j}^{(N)}\mathbf{%
\tilde{\eta}}_{sm,l}^{(k)}+\mathbf{f}_{l,j}^{(N)},  \label{eta_bp}
\end{equation}%
$\mathbf{W}_{1,l,j}^{(N)}[k]\triangleq (\mathbf{C}_{1,l,j}^{(N)}[k])^{-1}$
and%
\begin{equation}
\mathbf{C}_{1,l,j}^{(N)}[k]\triangleq \mathbf{A}_{l,j}^{(N)}\mathbf{\tilde{C}%
}_{sm,l}^{(k)}\left( \mathbf{A}_{l,j}^{(N)}\right) ^{T}+\mathbf{C}_{w}^{(N)}.
\label{cov_bp}
\end{equation}%
Then, the backward message $\overset{\leftarrow }{m}_{be1,j}^{(k)}(\mathbf{x}%
_{l}^{(N)})$ is evaluated as (see Fig. \ref{Fig_5}) 
\begin{equation}
\overset{\leftarrow }{m}_{be1,j}^{(k)}(\mathbf{x}_{l}^{(N)})=\overset{%
\leftarrow }{m}_{bp,j}^{(k)}(\mathbf{x}_{l}^{(N)})\,m_{pm,j}^{(k)}(\mathbf{x}%
_{l}^{(N)}).  \label{m_be1_k}
\end{equation}%
Based on eqs. (\ref{m_pm_x_N_l_j}) and (\ref{weight_bpa}), the last formula
can be rewritten as 
\begin{equation}
\overset{\leftarrow }{m}_{be1,j}^{(k)}(\mathbf{x}%
_{l}^{(N)})=w_{be1,l,j}^{(k)},  \label{m_be1_x_Na}
\end{equation}%
where 
\begin{equation}
w_{be1,l,j}^{(k)}\triangleq
w_{bp,l,j}^{(k)}\,w_{pm,l,j}^{(k)}\,=D_{be1,l,j}^{(k)}\exp \left( -\frac{1}{2%
}Z_{be1,l,j}^{(k)}\right)  \label{w_fe_2_x_N_l}
\end{equation}%
for any $j$, where $D_{be1,l,j}^{(k)}\triangleq
D_{pm,l,j}^{(k)}\,D_{bp,l,j}^{(k)}$ and $Z_{be1,l,j}^{(k)}\triangleq
Z_{pm,l,j}^{(k)}+Z_{bp,l,j}^{(k)}$. The messages $\{\overset{\leftarrow }{m}%
_{be1,j}^{(k)}(\mathbf{x}_{l}^{(N)})\}$ (i.e., the weights $%
\{w_{be1,l,j}^{(k)}\}$) are stored for the next iteration (see step1)).

Step 5) \emph{Computation of new measurement-based weights} in BIF$_{2}$-
The new measurement-based weight (see Fig. \ref{Fig_5}) 
\begin{eqnarray}
w_{fe1,l,j}^{(k)}\, &=&\int f(\mathbf{y}_{l}|\mathbf{x}_{fp,l,j}^{(N)},\,%
\mathbf{x}_{l}^{(L)})\,m_{sm}^{(k)}(\mathbf{x}_{l}^{(L)})\,d\mathbf{x}%
_{l}^{(L)}  \label{eq:mess_ms_PF} \\
&=&\mathcal{N}\left( \mathbf{y}_{l};\mathbf{\tilde{\eta}}_{ms,l,j}^{(k)},%
\mathbf{\tilde{C}}_{ms,l,j}^{(k)}\right) =m_{ms,j}^{(k)}\left( \mathbf{x}%
_{l}^{(N)}\right)  \notag \\
&&  \label{eq:weight_before_resampling}
\end{eqnarray}%
is computed on the basis of $m_{sm}^{(k)}(\mathbf{x}_{l}^{(L)})$ (\ref%
{m_fe_L_EKF_2}); here,%
\begin{equation}
\mathbf{\tilde{\eta}}_{ms,l,j}^{(k)}\triangleq \mathbf{B}_{l,j}\,\mathbf{%
\tilde{\eta}}_{sm,l}^{(k)}+\mathbf{g}_{l,j}  \label{eq:ms_fe_PF}
\end{equation}%
and%
\begin{equation}
\mathbf{\tilde{C}}_{ms,l,j}^{(k)}\triangleq \mathbf{B}_{l,j}\mathbf{\tilde{C}%
}_{sm,l}^{(k)}\mathbf{B}_{l,j}^{T}+\mathbf{C}_{e},  \label{eq:C_sm_PF}
\end{equation}%
where $\mathbf{B}_{l,j}\triangleq \mathbf{B}_{l}(\mathbf{x}_{fp,l,j}^{(N)})$
and $\mathbf{g}_{l,j}\triangleq \mathbf{g}_{l}(\mathbf{x}_{fp,l,j}^{(N)})$.
Then, the $N_{p}$ messages $\{m_{ms,j}^{(k)}(\mathbf{x}_{l}^{(N)})\}$ (i.e.,
the weights $\{w_{fe1,l,j}^{(k)}\,\}$) are stored, since in the next
iteration they are employed to generate the message (see Fig. \ref{Fig_5},
and eqs. (\ref{m_fp_N_MPF}) and (\ref{eq:weight_before_resampling}))%
\begin{eqnarray}
\overset{\rightarrow }{m}_{fe1,j}^{(k)}(\mathbf{x}_{l}^{(N)})
&=&m_{ms,j}^{(k)}\left( \mathbf{x}_{l}^{(N)}\right) \,\vec{m}_{fp,j}\left( 
\mathbf{x}_{l}^{(N)}\right)  \notag \\
&=&w_{fe1,l,j}^{(k)}\,\delta \left( \mathbf{x}_{l}^{(N)}-\mathbf{x}%
_{fp,l,j}^{(N)}\right)  \label{mess_N_fe1_j}
\end{eqnarray}%
and, then, the message $m_{sm,j}^{(k)}(\mathbf{x}_{l}^{(N)})$ (\ref%
{m_sm_j_N1}) (i.e., the smoothed weight $w_{sm,l,j}^{(k)}$ (\ref{w__sm_j_N}%
)); this concludes the $k$-th iteration. Then, the index $k$ is increased by
one, and a new iteration is started by going back to step 1) if $k<N_{it}+1$%
; otherwise (i.e., if $k=N_{it}+1$, we proceed with the next phase.

\textbf{Phase III} - In this phase, only step 1) and part of step 2) of
phase II are carried out in order to compute all the statistical information
required for the evaluation of the backward estimates $\overset{\leftarrow }{%
m}_{be}\left( \mathbf{x}_{l}\right) $ and $\overset{\leftarrow }{m}_{be}(%
\mathbf{x}_{l}^{(N)})$, i.e. the outputs generated by BIF$_{1}$ and BIF$_{2}$%
, respectively, in the $l$-th recursion of TS. More specifically, the
smoothed information $\{m_{sm,j}^{(N_{it}+1)}(\mathbf{x}_{l}^{(N)})\}$ is
computed (as if an additional iteration was started; see eqs. (\ref%
{m_sm_j_N2})-(\ref{w__sm_j_N})), the new weights $\{W_{sm,l,j}^{(N_{it}+1)}%
\} $ are evaluated on the basis of eq. (\ref{W_fe_2_x_N_l}) and the set $%
S_{fp,l}$ is sampled once on the basis of such weights; if the $j_{l}$-th
particle (i.e., $\mathbf{x}_{fp,l,j_{l}}^{(N)}$) is selected, we set 
\begin{equation}
\mathbf{x}_{be,l}^{(N)}=\mathbf{x}_{fp,l,j_{l}}^{(N)},
\label{selected_particle}
\end{equation}%
so that the message $\overset{\leftarrow }{m}_{be}(\mathbf{x}_{l}^{(N)})$ (%
\ref{mess_be_N_l}) becomes available at the output of BIF$_{1}$. On the
other hand, the evaluation of the message $\overset{\leftarrow }{m}%
_{be}\left( \mathbf{x}_{l}\right) $ is accomplished as follows. The messages 
$m_{pm}^{(N_{it}+1)}(\mathbf{x}_{l})$ and $\overset{\leftarrow }{m}%
_{be1,l}^{(N_{it}+1)}\left( \mathbf{x}_{l}\right) $ are computed first (see
eq. (\ref{m_pm_x_l}) and eqs. (\ref{m_be1_x_la})-(\ref{m_be1_x_laa}),
respectively). Then, the BIF$_{2}$ output message $\overset{\leftarrow }{m}%
_{be,l}\left( \mathbf{x}_{l}\right) $ is computed as (see Fig. \ref{Fig_5}) 
\begin{eqnarray}
\overset{\leftarrow }{m}_{be,l}\left( \mathbf{x}_{l}\right) &=&\overset{%
\leftarrow }{m}_{be2,l}\left( \mathbf{x}_{l}\right) =\overset{\leftarrow }{m}%
_{be1,l}^{(N_{it}+1)}\left( \mathbf{x}_{l}\right) \,m_{ms}\left( \mathbf{x}%
_{l}\right)  \label{m_be2_x} \\
&=&\mathcal{\mathcal{N}}\left( \mathbf{x}_{l};\mathbf{\eta }_{be2,l},\mathbf{%
C}_{be2,l}\right) ,  \label{m_be2_xb}
\end{eqnarray}%
where%
\begin{equation}
m_{ms}\left( \mathbf{x}_{l}\right) =\mathcal{\mathcal{N}}\left( \mathbf{x}%
_{l};\mathbf{\eta }_{ms,l},\mathbf{C}_{ms,l}\right)  \label{m_ms_x}
\end{equation}%
is the message conveying the measurement information. Moreover, the
covariance matrices $\mathbf{C}_{ms,l}$ and $\mathbf{C}_{be2,l}$, and the
mean vectors $\mathbf{\eta }_{ms,l}$ and $\mathbf{\eta }_{be2,l}$ are
computed on the basis of the associated precision matrices 
\begin{equation}
\mathbf{W}_{ms,l}\triangleq (\mathbf{C}_{ms,l})^{-1}=\mathbf{H}_{l}\mathbf{W}%
_{e}\mathbf{H}_{l}^{T},  \label{W_ms_x}
\end{equation}%
\begin{equation}
\mathbf{W}_{be2,l}\triangleq (\mathbf{C}_{be2,l})^{-1}=\mathbf{W}_{ms,l}+%
\mathbf{W}_{be1,l}^{(N_{it}+1)},  \label{W_be2_x}
\end{equation}%
and of the transformed mean vectors%
\begin{equation}
\mathbf{w}_{ms,l}\triangleq \mathbf{W}_{ms,l}\,\mathbf{\eta }_{ms,l}=\mathbf{%
H}_{l}\mathbf{W}_{e}\left( \mathbf{y}_{l}-\mathbf{v}_{l}\right) \text{,}
\label{w_ms_x}
\end{equation}%
\begin{equation}
\mathbf{w}_{be2,l}\triangleq \mathbf{W}_{be2,l}\mathbf{\eta }_{be2,l}=%
\mathbf{w}_{ms,l}+\mathbf{w}_{be1,l}^{(N_{it}+1)},  \label{w_be2_x}
\end{equation}%
respectively. The $l$-th recursion is now over.

It is important to point out that the \emph{first recursion} of the backward
pass requires the knowledge of the input messages $\overset{\leftarrow }{m}%
_{be}(\mathbf{x}_{T})$ and $\overset{\leftarrow }{m}_{be}(\mathbf{x}%
_{T}^{(N)})$. Similarly as any BIF algorithm, the evaluation of these
messages in BITF is based on the statistical information generated in the
last recursion of the forward pass. In particular, the above mentioned
messages are still expressed by eqs. (\ref{m_be2}) and (\ref{m_be2_N}) (with 
$l=T$ in both formulas), respectively. However, the vector $\mathbf{x}%
_{be,T}^{(N)}$ is generated by sampling the particle set $S_{fp,T}$ on the
basis of the forward weights $\{w_{fe,T,j}\}$, since backward predictions
are unavailable at the final instant $l=T$. Therefore, if the $j_{T}$-th
particle of $S_{fp,T}$ is selected, we set%
\begin{equation}
\mathbf{x}_{be,T}^{(N)}=\mathbf{x}_{fe,l,j_{T}}^{(N)}  \label{x_be_N_T}
\end{equation}%
in the message $\overset{\leftarrow }{m}_{be}(\mathbf{x}_{T}^{(N)})$
entering the BIF$_{2}$ in the first recursion (see eq. (\ref{mess_be_N_l})).
As far as BIF$_{1}$ is concerned, following \cite{Vitetta_2018}, we choose%
\begin{equation}
\mathbf{W}_{be,T}=\mathbf{W}_{fe1,T}  \label{C_be_L_T}
\end{equation}%
and%
\begin{equation}
\mathbf{w}_{be,T}=\mathbf{w}_{fe1,T}  \label{eta_be_L_T}
\end{equation}%
for the message $\overset{\leftarrow }{m}_{be}(\mathbf{x}_{T})$.

The general method for BITF and TS developed in this Paragraph is summarized
in Algorithm 1.

\begin{algorithm}{
\SetKw{a}{a-}
\SetKw{b}{b-}
\SetKw{c}{c-}%

\SetKw{d}{d-}
\SetKw{e}{e-}
\SetKw{f}{f-}
\SetKw{g}{g-}
\SetKw{h}{h-}%

\nl\textbf{Forward filtering}: For $l=1$ to $T$: Run a TF\ algorithm,
and store $\mathbf{W}_{fe1,l}$ (\ref{W_fe_L}), $\mathbf{w}_{fe1,l}$ (\ref{w_fe_L}), $S_{fp,l}=\{\mathbf{x}%
_{fp,l,j}^{(N)}\}$ and $\{{w}_{fe,l,j}\}_{j=1}^{N_{p}}$.

\nl\textbf{%
Initialisation of backward filtering}: compute $\mathbf{x}_{be,T}^{(N)}$ (%
\ref{x_be_N_T}), $\mathbf{W}%
_{be,T}$ (\ref{C_be_L_T}) and $\mathbf{w}_{be,T}$ (\ref{eta_be_L_T}); then, compute $\mathbf{C}_{be,T}=(\mathbf{W}%
_{be,T})^{-1}$, $\mathbf{\eta}_{be,T}=\mathbf{C}_{be,T}\mathbf{w}_{be,T}$.

\nl\textbf{Backward filtering and smoothing}: \\
\For {$l=T-1$ to $1$}{
	\a \textbf{Phase I}:
	
		- \emph{Marginalization}: extract $\mathbf{%
\tilde{\eta}}_{be,l+1}$ ($\mathbf{\tilde{C}}_{be,l+1}$) from $\mathbf{\eta}%
_{be,l+1}$ ($\mathbf{C}_{be,l+1}$).

		- \emph{Backward filter prediction}%
: compute $\mathbf{W}_{bp,l}$ (\ref{W_bp_x_l}) and
 $\mathbf{w}_{bp,l}$ (%
\ref{w_bp_x_l}).
		
		- \emph{Computation of the pseudo-measurements for}
BIF$_1$: For $j=1$ to $N_{p}$: compute $\mathbf{z}_{l,j}^{(L)}$ (\ref{PM_z_L}%
), $\mathbf{\tilde{W}}_{pm,l,j}$ (\ref{eq:W_pm_L_j}), $\mathbf{\tilde{w}}%
_{pm,l,j}$ (%
\ref{eq:w_pm_L_j}), $\mathbf{\tilde{C}}_{pm,l,j}=(\mathbf{%
\tilde{W}}_{pm,l,j})^{-1}$ and $\mathbf{\tilde{\eta}}_{pm,l,j}=\mathbf{%
\tilde{C}}_{pm,l,j}\mathbf{\tilde{w}}_{pm,l,j}$.

    - \emph{%
Initialisation of particle weights}: Set  $w_{be1,l,j}^{(0)}=1$ and $w_{fe1,l,j}^{(0)}=w_{fe,l,j}$.\\
		
	 \textbf{Phase II}:\\
	\For {$k=1$ to $N_{it}$}{
		
			\b
Step 1): For $j=1$ to $N_{p}$: Compute $w_{sm,l,j}^{(k)}$ (\ref{w__sm_j_N})
and $W_{sm,l,j}^{(k)}$ (\ref{W_fe_2_x_N_l}); then, compute 
$\mathbf{\eta}%
_{pm,l}^{(k)}$ (\ref{eta_pm_l_k}) and $\mathbf{C}_{pm,l}^{(k)}$ (\ref{C_pm_l_k}).
			
			\c Step 2): compute $\mathbf{C}_{be1,l}^{(k)}$ (\ref{C_be1_l_ka}), $\mathbf{\eta }_{be1,l}^{(k)}$ (\ref{eta_be1_l_ka}), $\mathbf{W%
}_{be1,l}^{(k)}=(\mathbf{C}_{be1,l}^{(k)})^{-1}$
			and $\mathbf{w}%
_{be1,l}^{(k)}=\mathbf{W}_{be1,l}^{(k)}\mathbf{\eta }_{be1,l}^{(k)}$, $%
\mathbf{W}_{sm,l}^{(k)}$ (\ref{W_sm_l_k}), $\mathbf{w}_{sm,l}^{(k)}$ (\ref{w_sm_l_k}), $\mathbf{C}_{sm,l}^{(k)}=(\mathbf{C}_{sm,l}^{(k)})^{-1}$ and $%
\mathbf{\eta}_{sm,l}^{(k)}=\mathbf{C}_{sm,l}^{(k)}\mathbf{w}_{sm,l}^{(k)}$. Then, extract $\mathbf{\tilde{\eta}}_{sm,l}^{(k)}$ ($\mathbf{\tilde{C}}%
_{sm,l}^{(k)}$) from $\mathbf{\eta}_{sm,l}^{(k)}$ ($\mathbf{C}_{sm,l}^{(k)}$%
).
			
			\d Step 3): For $j=1$ to $N_{p}$: compute $\mathbf{\eta}%
_{z,l,j}^{(k)}$ (\ref{eta_mess_z_N}), $\mathbf{C}_{z,l,j}^{(k)}$ (\ref{C_mess_Z_N_bis}), $\mathbf{W}_{z,l,j}^{(k)}=(\mathbf{C}_{z,l,j}^{(k)})^{-1}$%
, $\mathbf{w}_{z,l,j}^{(k)}=\mathbf{W}_{z,l,j}^{(k)}\mathbf{\eta}%
_{z,l,j}^{(k)}$, $\mathbf{W}_{pm,l,j}^{(k)}$ (\ref{W_pm_x_N_l_j}), $\mathbf{w%
}_{pm,l,j}^{(k)}$ (\ref{w_pm_x_N_l_j}) and $w_{pm,l,j}^{(k)}$ (\ref{m_pm_x_N_l_j}).
			  	
			\e Step 4): For $j=1$ to $N_{p}$: compute $%
\mathbf{\eta}_{1,l,j}^{(N)}[k]$ (\ref{eta_bp}), $\mathbf{C}_{1,l,j}^{(N)}[k]$
(\ref{cov_bp}), $w_{bp,l,j}^{(k)}$ (\ref{weight_bpa}) and $w_{be1,l,j}^{(k)}$
(\ref{w_fe_2_x_N_l}).
			
		\f Step 5): For $j=1$ to $N_{p}$: Compute $%
\mathbf{\tilde{\eta}}_{ms,l,j}^{(k)}$ (\ref{eq:ms_fe_PF}), $\mathbf{\tilde{C}%
}_{ms,l,j}^{(k)}$ (\ref{eq:C_sm_PF}) and $w_{fe1,l,j}^{(k)}$ (\ref{%
eq:weight_before_resampling}).
			

}

\g \textbf{Phase III} - BIF$%
_2$: set $k=N_{it}+1$ and compute 
the new particle weights $%
\{W_{sm,l,j}^{(N_{it}+1)}\}$ (see step 1)). Then, select the $j_{l}$-th
particle 
$\mathbf{x}_{fp,l,j_{l}}^{(N)}$ by sampling the set $S_{fp,l}$ on
the basis of these weights, 
set $\mathbf{x}_{be,l}^{(N)}=\mathbf{x}%
_{fp,l,j_{l}}^{(N)}$ and store $\mathbf{x}_{be,l}^{(N)}$ 
for the next
recursion.

 \h \textbf{Phase III} - BIF$_1$: 
Compute $\mathbf{\eta}%
_{pm,l}^{(N_{it}+1)}$,  $\mathbf{C}_{pm,l}^{(N_{it}+1)}$, $\mathbf{W}%
_{be1,l}^{(N_{it}+1)}$  and $\mathbf{w}_{be1,l}^{(N_{it}+1)}$  (see step
1)). 
Then, compute  $\mathbf{W}_{ms,l}$ (\ref{W_ms_x}), $\mathbf{w}_{ms,l}$
(\ref{w_ms_x}), $\mathbf{W}_{be2,l}$ (\ref{W_be2_x}), 
$\mathbf{w}_{be2,l}$
(\ref{w_be2_x}), $\mathbf{C}_{be,l}=(\mathbf{W}_{be2,l})^{-1}$ and
 $%
\mathbf{\eta}_{be,l}=\mathbf{C}_{be,l}\mathbf{w}_{be2,l}$, and store $\mathbf{C}_{be,l}$ and $\mathbf{\eta}_{be,l}$ for the next recursion.
}
\caption{Backward Information Turbo Filtering and Turbo Smoothing}}
\end{algorithm}

Algorithm 1 produces all the statistical information required to solve
problems \textbf{P.1} and \textbf{P.2}. Let us now discuss how this can be
done in detail. As far as problem\textbf{\ P.1} is concerned, it is useful
to point out that Algorithm 1 produces a trajectory $\{\mathbf{x}%
_{be,l}^{(N)},l=1,2,...,T\}$ for the \emph{nonlinear} component (see eq. (%
\ref{selected_particle})). Another trajectory, representing the time
evolution of the \emph{linear} state component only and denoted $\{\mathbf{x}%
_{be,l}^{(L)},l=1,2,...,T\}$, can be computed by sampling the message $\vec{m%
}_{sm}^{(N_{it}+1)}(\mathbf{x}_{l}^{(L)})$ (see eq. (\ref{m_fe_L_EKF_2})) or
by simply setting $\mathbf{x}_{be,l}^{(L)}=\mathbf{\tilde{\eta}}%
_{sm,l}^{(N_{it}+1)}$ (this task can be accomplished in task in step 3-h of
Algorithm 1, after sampling the particle set $S_{fp,l}$). The overall
algorithm producing this result is called \emph{turbo smoothing algorithm }%
(TSA) in the following.

The TSA solves problem \textbf{P.1} and, consequently, problem \textbf{P.2},
since, once it has been run, an approximation of the marginal smoothed pdf
at any instant can be simply obtained by marginalization. The last result,
however, is achieved at the price of a significant computational cost since $%
M$ backward passes are required. However, if we are interested in solving
problem P.2 only, a simpler particle smoother can be developed following the
approach illustrated in \cite{Vitetta_2018}, so that a single backward pass
has to be run. In this pass, the evaluation of the message $\overset{%
\leftarrow }{m}_{be}(\mathbf{x}_{l}^{(N)})$ (i.e., of the particle $\mathbf{x%
}_{be,l}^{(N)}$) involves the whole particle set $S_{fp,l}$ and their
weights $\{W_{sm,l,j}^{(N_{it}+1)}\}$ (see eq. (\ref{W_fe_2_x_N_l}))
evaluated in the last phase of the $l$-th recursion. More specifically, a
new smoother is obtained by employing a different method for evaluating $%
\mathbf{x}_{be,l}^{(N)}$ in step 3-h of Algorithm 1; it consists in
computing the smoothed estimate%
\begin{equation}
\mathbf{x}_{sm,l}^{(N)}=\sum\limits_{j=0}^{N_{p}-1}W_{sm,l,j}^{(N_{it}+1)}\,%
\mathbf{x}_{fp,l,j}^{(N)}  \label{x_sm_N}
\end{equation}%
of $\mathbf{x}_{l}^{(N)}$ and, then, setting%
\begin{equation}
\mathbf{x}_{be,l}^{(N)}=\mathbf{x}_{sm,l}^{(N)}.  \label{x_be_N_new}
\end{equation}%
The resulting smoother is called \emph{simplified} \emph{turbo smoothing
algorithm} (STSA) in the following.

Finally, it is important to point out that the computational complexity of
the TSA and the STSA can be substantially reduced by reusing the forward
weights $\{w_{fe1,l,j}\}$ in all the iterations of phase II, so that step 5)
can be skipped; this means that, for any $k$, we set $%
w_{fe1,l,j}^{(k-1)}=w_{fe1,l,j}$ in the evaluation of the $j$-th particle
weight $w_{sm,l,j}^{(k)}$ according to eq. (\ref{w__sm_j_N}) in step 1). Our
simulation results have evidenced that, at least for the SSM considered in
Section \ref{num_results}, this modification does not have any impact on the
estimation accuracy of these algorithms.

\subsection{Comparison of the Developed Turbo Smoothing Algorithms with
Related Techniques\label{Comparison}}

The TSA developed in the previous Section is conceptually related to the 
\emph{Rao-Blackwellized particle smoothing} (RBPS) techniques proposed by
Fong \emph{et al. }\cite{Fong_2002} and by Lindsten \emph{et al.} \cite%
{Lindsten_2016} (these algorithms are denoted Alg-B and Alg-L respectively,
in the following) and to the RBSS algorithm devised by Vitetta \emph{et al.} 
\cite{Vitetta_2018}. In fact, all these techniques share with the TSA the
following important features: 1) all of them aim at estimating the \emph{%
joint} smoothing density over the whole observation interval by generating
multiple \emph{realizations} from it; 2) they accomplish a single forward
pass and as many backward passes as the overall number of realizations; 3)\
they combine Kalman filtering with particle filtering. However, Alg-B, Alg-L
and the RBSS\ algorithm employ, in both their forward and backward passes,
as many Kalman filters as the number of particles ($N_{p}$) to generate a
particle-dependent estimate of the linear state component only. On the
contrary, the TSA employs a single (extended) Kalman filter, that, however,
estimates the whole system state. This substantially reduces the memory
requirements of particle smooothing and, consequently, the overall number of
memory accesses accomplished on the hardware platform on smoothing is run;
as evidenced by our numerical results, this feature contributes to making
the overall execution time of TSA appreciably shorter than that required by
the related algorithms.

On the other hand, the STSA is conceptually related to the SPS algorithm
devised by Vitetta \emph{et al.} \cite{Vitetta_2018}. In fact, both
algorithms aim at solving problem \textbf{P.2} only and, consequently, carry
out a \emph{single backward pass}. This property makes them much faster than
Alg-B, Alg-L and the RBSS algorithm in the computation of marginal smoothed
densities. Finally, note that, similarly as the TS technique, the use of the
STSA requires a substantially smaller number of memory accesses than the SPS
algorithm.

\section{Numerical Results\label{num_results}}

In this Section we compare, in terms of accuracy and execution time, the TSA
and the STSA with Alg-L, the RBSS and the SPS algorithm for a specific CLG
SSM. The considered SSM is the same as the SSM\#2 defined in \cite%
{Vitetta_2018} and describes the bidimensional motion of an agent. Its state
vector in the $l$-th observation interval is defined as $\mathbf{x}%
_{l}\triangleq \lbrack \mathbf{v}_{l}^{T},\mathbf{p}_{l}^{T}]^{T}$, where $%
\mathbf{v}_{l}\triangleq \lbrack v_{x,l},v_{y,l}]^{T}$ and $\mathbf{p}%
_{l}\triangleq \lbrack p_{x,l},p_{y,l}]^{T}$ (corresponding to $\mathbf{x}%
_{l}^{(L)}$ and $\mathbf{x}_{l}^{(N)}$, respectively) represent the agent
velocity and position, respectively (their components are expressed in m/s
and in m, respectively). The state update equations are%
\begin{equation}
\mathbf{v}_{l+1}=\rho \mathbf{v}_{l}+T_{s}\mathbf{a}_{l}(\mathbf{p}%
_{l})+\left( 1-\rho \right) \mathbf{n}_{v,l}  \label{mod_1_v}
\end{equation}%
and%
\begin{equation}
\mathbf{p}_{l+1}=\mathbf{p}_{l}+\mathbf{v}_{l}\cdot T_{s}+(T^{2}/2)\mathbf{a}%
_{l}(\mathbf{p}_{l})+\mathbf{n}_{p,l},  \label{mod_1_p}
\end{equation}%
where $\rho $ is a forgetting factor (with $0<\rho <1$), $T_{s}$ is the
sampling interval, $\mathbf{n}_{v,l}$ is an \emph{additive Gaussian noise}
(AGN) vector characterized by the covariance matrix $\mathbf{I}_{2}$, 
\begin{equation}
\mathbf{a}_{l}\left( \mathbf{p}_{l}\right) =-a_{0}\frac{\mathbf{p}_{l}}{%
\left\Vert \mathbf{p}_{l}\right\Vert }\frac{1}{1+\left( \left\Vert \mathbf{p}%
_{l}\right\Vert /d_{0}\right) ^{2}}  \label{acc_model}
\end{equation}%
is the acceleration due to a force applied to the agent (and pointing
towards the origin of our reference system), $a_{0}$ is a scale factor
(expressed in m/s$^{2}$), $d_{0}$ is a \emph{reference distance} (expressed
in m), and $\mathbf{n}_{p,l}$ is an AGN vector characterized by the
covariance matrix $\sigma _{p}^{2}\mathbf{I}_{2}$ and accounting for model
inaccuracy. The measurement vector available in the $l$-th interval for
state estimation is 
\begin{equation}
\mathbf{y}_{l}=\mathbf{x}_{l}+\mathbf{e}_{l},  \label{mod_1_y}
\end{equation}%
where $\mathbf{e}_{l}\triangleq \lbrack \mathbf{e}_{v,l}^{T},\mathbf{e}%
_{p,l}^{T}]^{T}$ and $\mathbf{e}_{v,l}$ ($\mathbf{e}_{p,l}$) is an AGN
vector characterized by the covariance matrix $\sigma _{ev}^{2}\mathbf{I}%
_{2} $ ($\sigma _{ep}^{2}\mathbf{I}_{2}$).

In our computer simulations, following \cite{Vitetta_2018} and \cite%
{Vitetta_2019}, the estimation accuracy of the considered smoothing
techniques has been assessed by evaluating two \emph{root mean square errors}
(RMSEs), one for the linear state component, the other for the nonlinear
one, over an observation interval lasting $T=200$ $T_{s}$; these are denoted 
$RMSE_{L}($alg$)$ and $RMSE_{N}($alg$)$, respectively, where `alg' is the
acronym of the algorithm these parameters refer to. Our assessment of
computational requirements is based, instead, on assessing the average \emph{%
computation time} required for processing a single \emph{block} of
measurements (this quantity is denoted CTB$($alg$)$ in the following).
Moreover, the following values have been selected for the parameters of the
considered SSM: $\rho =0.995$, $T_{s}=0.01$ s, $\sigma _{p}$ $=5\cdot
10^{-3} $ m, $\sigma _{e,p}=2\cdot 10^{-2}$ m, $\sigma _{e,v}=2\cdot 10^{-2}$
m/s, $a_{0}=0.5$ m/s$^{2}$, $d_{0}=5\cdot 10^{-3}$ m and $v_{0}=1$ m/s (the
initial position $\mathbf{p}_{0}\triangleq \lbrack p_{x,0},p_{y,0}]^{T}$ and
the initial velocity $\mathbf{v}_{0}\triangleq \lbrack v_{x,0},v_{y,0}]^{T}$
have been set to $[0.01$ m$,$ $0.01$ m$]^{T}$ and $[0.01$ m/s$,$ $0.01$ m/s$%
]^{T}$, respectively).

\begin{figure}[tbp]
\centering
\includegraphics[width=0.70\textwidth]{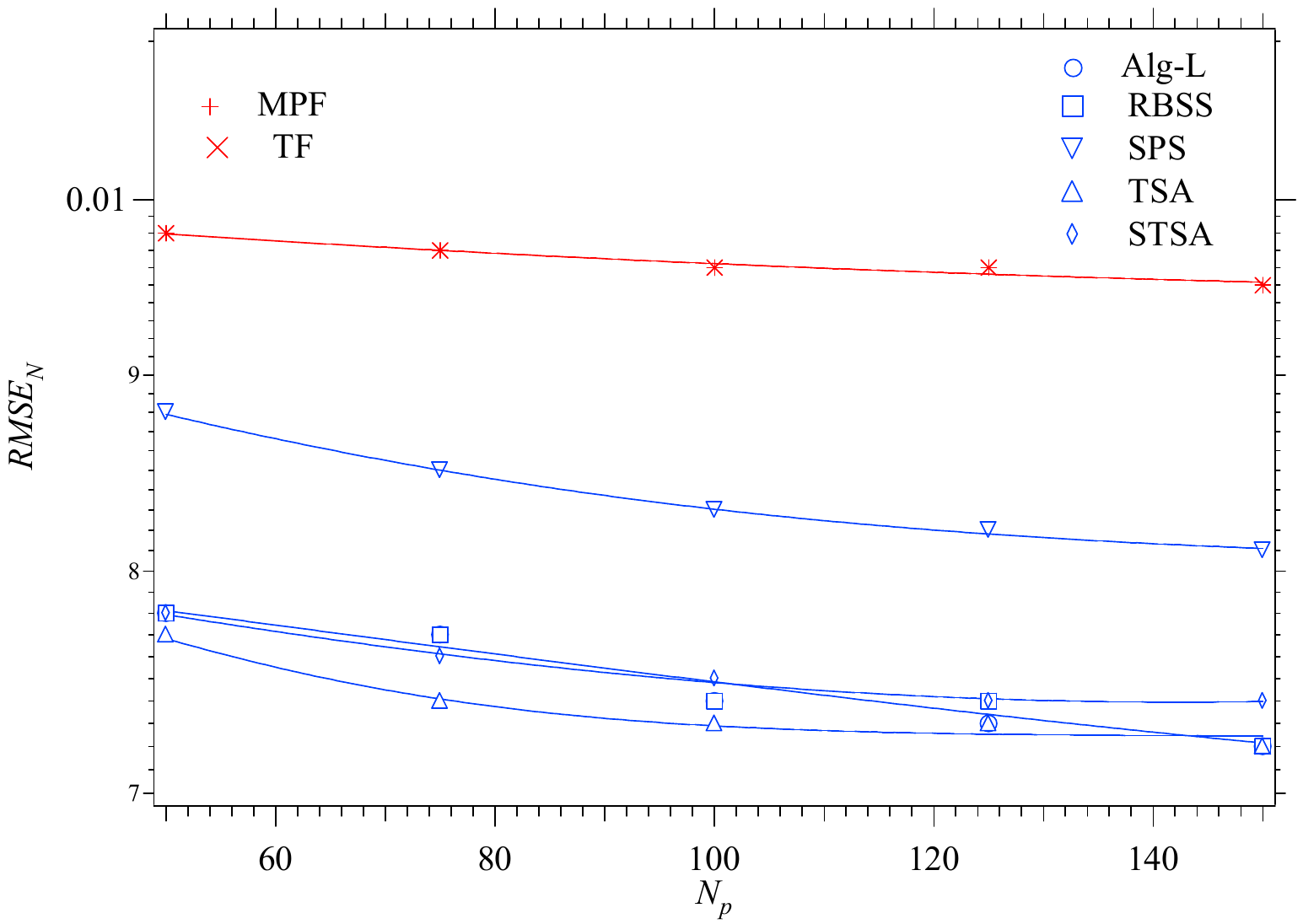}
\caption{RMSE performance versus $N_{p}$ for the nonlinear component ($%
RMSE_{N}$) of system state; five smoothing algorithms (Alg-L, the TSA, the
STSA, and the RBSS and SPS algorithms) and two filtering techniques (MPF and
TF) are considered.}
\label{Fig_rmsen}
\end{figure}

\begin{figure}[tbp]
\centering
\includegraphics[width=0.70\textwidth]{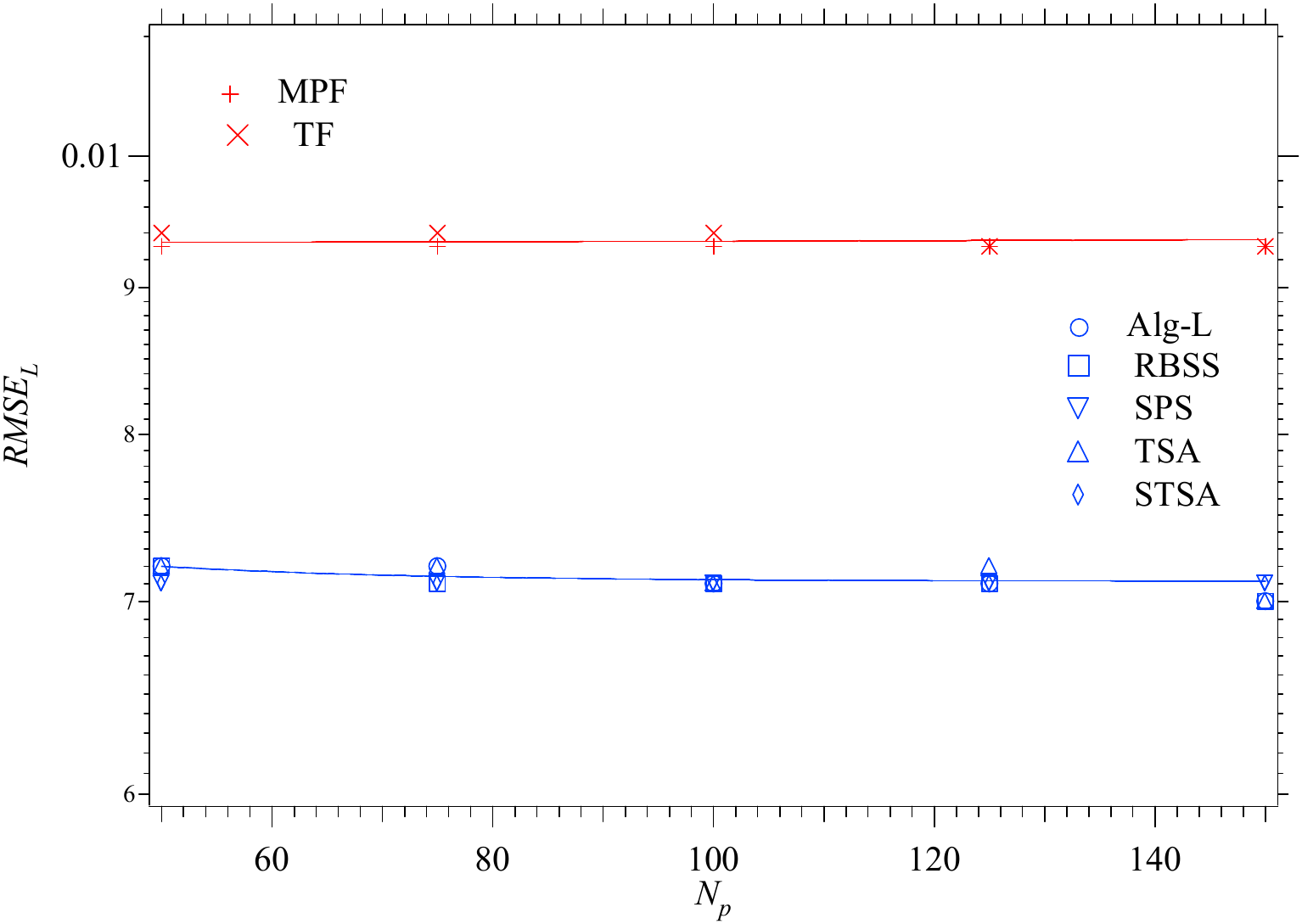}
\caption{RMSE performance versus $N_{p}$ for the linear component ($RMSE_{L}$%
) of system state; five smoothing algorithms five smoothing algorithms
(Alg-L, the TSA, the STSA, and the RBSS and SPS algorithms) and two
filtering techniques (MPF and TF) are considered.}
\label{Fig_rmsel}
\end{figure}

Some numerical results showing the dependence of $RMSE_{L}$ and $RMSE_{N}$
on the number of particles ($N_{p}$) for the considered smoothing algorithms
are illustrated in Figs. \ref{Fig_rmsen} and \ref{Fig_rmsel}, respectively
(simulation results are indicated by markers, whereas continuous lines are
drawn to fit them, so facilitating the interpretation of the available
data). In this case, $N_{it}=1$ has been selected for both the TSA and the
STSA, and the range $[10,150]$ has been considered for $N_{p}$ (since no
real improvement is found for $N_{p}\gtrsim 150$). Morever, $RMSE_{L}$ and $%
RMSE_{N}$ results are also provided for MPF (TF with $N_{it}=1$), since this
filtering technique is employed in the forward pass of Alg-L, the RBSS
algorithm and the SPS algorithm (the TSA and the STSA); this allows us to
assess the improvement in estimation accuracy provided by the backward pass
with respect to the forward pass for each smoothing algorithm. These results
show that:

1) The TSA, the STSA, Alg-L and the RBSS algorithm achieve similar
accuracies in the estimation of both the linear and nonlinear state
components.

2) The SPS algorithm is slightly outperformed by the other four smoothing
algorithms in terms of $RMSE_{N}$ only; for instance, $RMSE_{N}($SPS$)$ is
about $1.11$ times larger than $RMSE_{N}($STSA$)$ for $N_{p}=100$.

3) Even if the RBSS algorithm and the TSA provide by far richer statistical
information than their simplified counterparts (i.e., than the SPS algorithm
and the STSA, respectively), they do not provide a significant improvement
in the accuracy of state estimation; for instance, $RMSE_{N}($SPS$)$ ($%
RMSE_{N}($STSA$)$) is about $1.12$ ($1.03$) time larger than $RMSE_{N}($RBSS$%
)$ ($RMSE_{N}($TSA$)$) for $N_{p}=100$.

4) The accuracy improvement in terms of $RMSE_{L}$ ($RMSE_{N}$) provided by
all the smoothing algorithms except the SPS (Alg-L, RBSS, TSA and the STSA)
is about $24\%$ (roughly $23\%$) with respect to the MPF and TF techniques,
for $N_{p}=100$. Moreover, the accuracy improvement in terms of $RMSE_{L}$ ($%
RMSE_{N}$) achieved by the SPS algorithm is about $24\%$ (about $14\%$) with
respect to the MPF technique for $N_{p}=100$.

Note also that, in the considered scenario, TF is slightly outperformed by
(perform similarly as) MPF in the estimation of the linear (nonlinear) state
component; a similar result is reported in \cite{Vitetta_2018_TF} for a
different SSM.

\begin{figure}[tbp]
\centering
\includegraphics[width=0.70\textwidth]{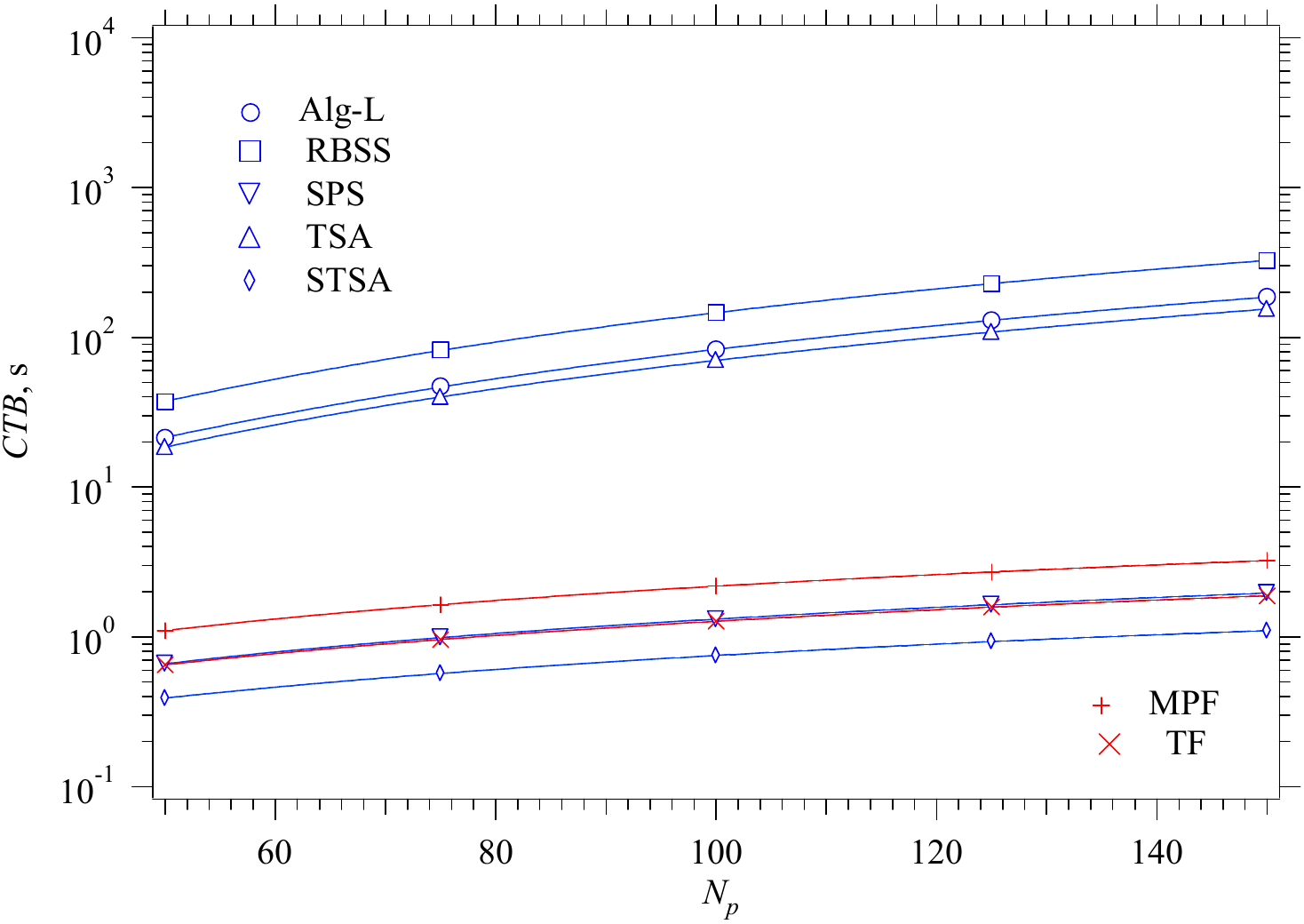}
\caption{CTB versus $N_{p}$ for five smoothing algorithms (Alg-L, TSA, STSA
and the RBSS and SPS algorithms) and two filtering techniques (MPF and TF).}
\label{Fig_CTB}
\end{figure}

Despite their similar accuracies, the considered smoothing algorithms
require different computational efforts; this is easily inferred from the
numerical results appearing in Fig. \ref{Fig_CTB} and illustrating the
dependence of the CTB on $N_{p}$ for all the above mentioned filtering and
smoothing algorithms. In fact, these results show that the TSA requires a
shorter computation time than Alg-L and the RBSS algorithm; more
specifically, CTB$($TSA$)$ is approximately $0.85$ ($0.48$) times smaller
than CTB$($Alg-L$)$ (CTB$($RBSS$)$). The same considerations apply to the
STSA and the SPS algorithm; in fact, CTB$($STSA$)$ is approximately $0.57$
times smaller than CTB$($SPS$)$. Note also that CTB$($TF$)$ is approximately 
$0.55$ times smaller than CTB$($MPF$)$ for the same value of $N_{p}$; once
again, this result is in agreement with the results shown in \cite%
{Vitetta_2018_TF} for a different SSM.

Finally, all the numerical results illustrated above lead to the conclusion
that, in the considered scenario, the TSA and STSA achieve the best
accuracy-complexity tradeoff in their categories of smoothing techniques.

\section{Conclusions\label{sec:conc}}

In this manuscript, factor graph methods have been exploited to formalise
the concept of parallel concatenation of Bayesian information filters. This
has allowed us to develop a new approximate method for Bayesian smoothing,
called \emph{turbo smoothing}. Two turbo smoothers have been derived for the
class of CLG systems and have been compared, in terms of both accuracy and
execution time, with other smoothing algorithms for a specific dynamic
model. These smoothers have limited requirements in terms of memory;
moreover, our simulation results evidence that they perform similarly as
their counterparts, but are faster.

\section*{Appendix}

In this Appendix, the derivation of the expressions of various messages
evaluated in each of the three phases the TFA consists of is sketched.

\textbf{Phase I} - Formulas (\ref{W_bp_x_l}) and (\ref{w_bp_x_l}), referring
to the message $\overset{\leftarrow }{m}_{bp}(\mathbf{x}_{l})$ (\ref{m_bp}),
can be easily computed by applying eqs. (IV.6)-(IV.8) of \ \cite[Table 4,
p.1304]{Loeliger_2007} in their backward form (with $A\rightarrow \mathbf{I}%
_{D}$, $X\rightarrow \mathbf{F}_{l}\mathbf{x}_{l}$, $Z\rightarrow \mathbf{x}%
_{l+1}$ and $Y\rightarrow \mathbf{u}_{l}+\mathbf{w}_{l}$) and, then, eqs.
(III.5)-(III.6) of \cite[Table 3, p.1304]{Loeliger_2007} (with $A\rightarrow 
\mathbf{F}_{l}$, $X\rightarrow \mathbf{x}_{l}$ and $Y\rightarrow \mathbf{F}%
_{l}\mathbf{x}_{l}$).

\begin{figure}[tbp]
\centering
\includegraphics[width=0.70\textwidth]{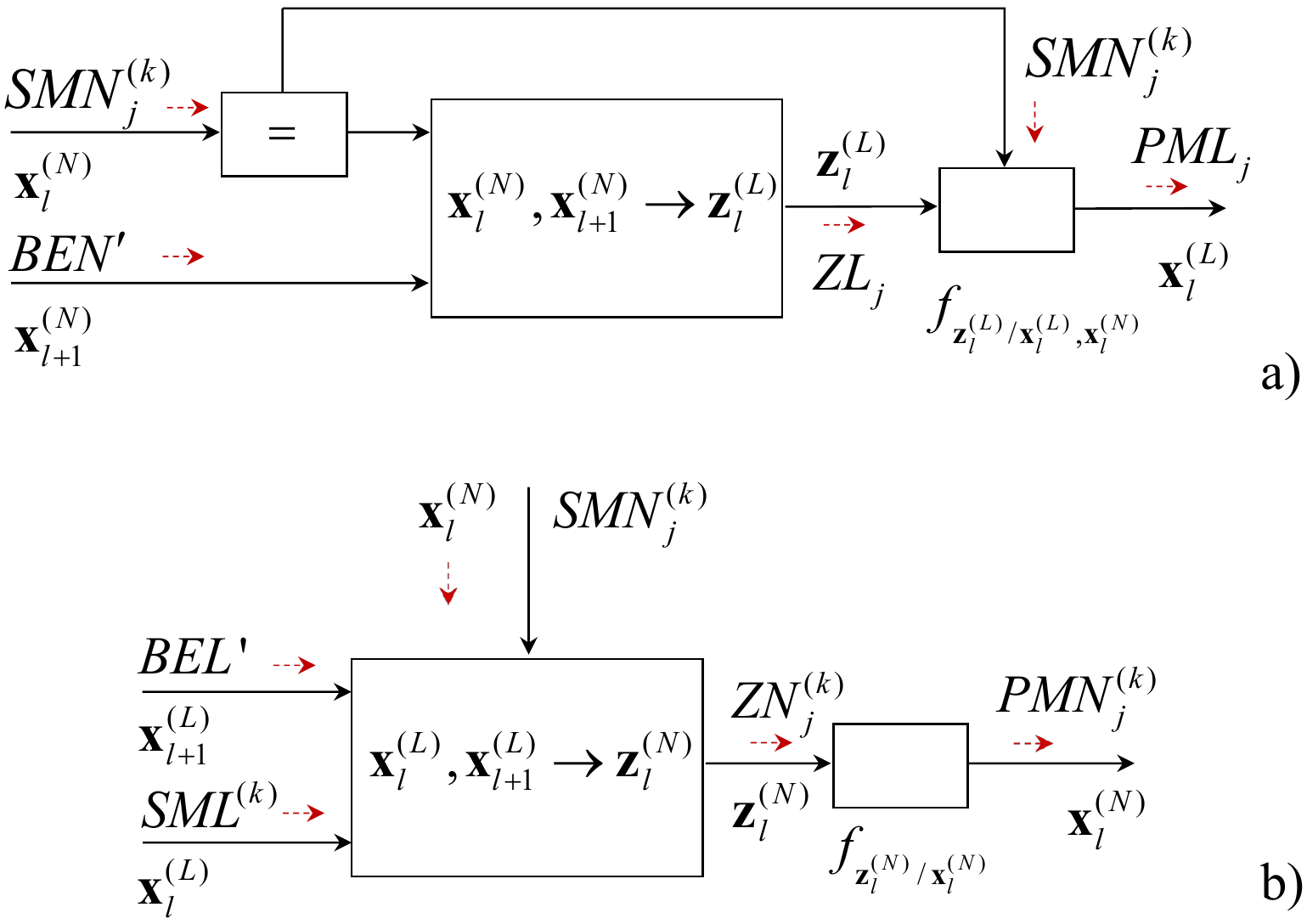}
\caption{Representation of the processing accomplished by a) the PMG$%
_{2\rightarrow 1}$ block and b) the PMG$_{1\rightarrow 2}$ block as message
passing over a factor graph.}
\label{Fig_6}
\end{figure}

The message set $\{m_{pm,j}(\mathbf{x}_{l}^{(L)})\}$ (see eq. (\ref%
{eq:message_pm_L_j_tris})) conveys the statistical information provided by
the pseudo-measurement\ $\mathbf{z}_{l}^{(L)}$ (\ref{eq:z_L_l}). The method
for computing the message $m_{pm,j}(\mathbf{x}_{l}^{(L)})$ can be
represented as a message passing over the graphical model shown in Fig. \ref%
{Fig_6}-a). Given $\mathbf{x}_{l}^{(N)}=\mathbf{x}_{fp,l,j}^{(N)}$ (this
particle is provided by the message $m_{sm,j}^{(k)}(\mathbf{x}_{l}^{(N)})$ (%
\ref{m_sm_j_N2})) and $\overset{\leftarrow }{m}_{be}(\mathbf{x}_{l+1}^{(N)})$
(\ref{mess_be_N_l}), the pseudo-measurement $\mathbf{z}_{l,j}^{(L)}$ (\ref%
{PM_z_L}) associated with the couple $(\mathbf{x}_{fp,l,j}^{(N)}$ , $\mathbf{%
x}_{be,l+1}^{(N)})$ is computed on the basis of eq. (\ref{eq:z_L_l}); this
pseudo-measurement is conveyed by the message (denoted $ZL_{j}$ in Fig. \ref%
{Fig_6}-a))%
\begin{equation}
m_{j}\left( \mathbf{z}_{l}^{(L)}\right) =\delta \left( \mathbf{z}_{l}^{(L)}-%
\mathbf{z}_{l,j}^{(L)}\right) ,  \label{m_j_z_L_bis}
\end{equation}%
which is employed in the evaluation of the message (see Fig. \ref{Fig_6}%
-(a)) 
\begin{equation}
m_{pm,j}\left( \mathbf{x}_{l}^{(L)}\right) =\int m_{j}\left( \mathbf{z}%
_{l}^{(L)}\right) \,f\left( \mathbf{z}_{l}^{(L)}\left\vert \mathbf{x}%
_{l}^{(L)},\mathbf{x}_{fp,l,j}^{(N)}\right. \right) d\mathbf{z}_{l}^{(L)}.
\label{eq:message_pm_L_j}
\end{equation}%
Then, substituting eq. (\ref{m_j_z_L_bis}) and $f(\mathbf{z}_{l}^{(L)}|%
\mathbf{x}_{l}^{(L)},\mathbf{x}_{l}^{(N)})=\mathcal{N(}\mathbf{z}_{l}^{(L)}$%
; $\mathbf{A}_{l,j}^{(N)}\mathbf{x}_{l}^{(L)},\mathbf{C}_{w}^{(N)})$ (see
eq. (\ref{eq:z_L_l_bis})) in the RHS of eq. (\ref{eq:message_pm_L_j}) yields
the message $m_{pm,j}^{(k)}(\mathbf{x}_{l}^{(L)})=\mathcal{\mathcal{N(}}%
\mathbf{z}_{l,j}^{(L)};\mathbf{A}_{l,j}^{(N)}\mathbf{x}_{l}^{(L)},\mathbf{C}%
_{w}^{(N)})$ (see \cite[App. A, TABLE\ II, formula no. 3]{Vitetta_2018}),
that can be easily put in the equivalent Gaussian form (\ref%
{eq:message_pm_L_j_tris}).

\textbf{Phase II} - Step 1) The procedure we adopt for computing $\overset{%
\leftarrow }{m}_{pm}^{(k)}(\mathbf{x}_{l})$ (\ref{m_pm_x_l}) on the basis of
the sets $\{m_{pm,j}(\mathbf{x}_{l}^{(L)})\}$ and $\{m_{sm,j}^{(k)}(\mathbf{x%
}_{l}^{(N)})\}$ (see eqs. (\ref{eq:message_pm_L_j_tris}) and (\ref{m_sm_j_N2}%
), respectively) is based on the following considerations. The message $%
m_{pm,j}(\mathbf{x}_{l}^{(L)})$ is \emph{coupled} with $m_{sm,j}^{(k)}(%
\mathbf{x}_{l}^{(N)})$ (for any $j$), since they refer to the same particle
set (i.e., $S_{fp,l}$). Moreover, these two messages provide \emph{%
complementary} information, because they refer to the two different
components of the overall state $\mathbf{x}_{l}$. For these reasons, the
statistical information conveyed by the above mentioned sets can be
condensed in the joint pdf%
\begin{equation}
f^{(k)}\left( \mathbf{x}_{l}^{(L)},\mathbf{x}_{l}^{(N)}\right) \triangleq
\sum\limits_{l=0}^{N_{p}-1}m_{sm,j}^{(k)}\left( \mathbf{x}_{l}^{(N)}\right)
m_{pm,j}(\mathbf{x}_{l}^{(L)}).  \label{joint_pdf}
\end{equation}%
Then, the message $m_{pm}^{(k)}(\mathbf{x}_{l})$ (\ref{m_pm_x_l}) can be
computed by projecting this pdf\emph{\ }onto a single Gaussian pdf; the
transformation adopted here to achieve this result and expressed by eqs. (%
\ref{eta_pm_l_k})-(\ref{C_pm_l_L_k_bis}) is described in \cite[Sec. IV]%
{Runnalls_2007}, and ensures that the \emph{mean }and the\emph{\ covariance}
of the given pdf are preserved.

Step 2) The expression (\ref{m_be1_x_laa}) of $\overset{\leftarrow }{m}%
_{be1}^{(k)}(\mathbf{x}_{l})$ represents a straightforward application of
formula no. 2 of \cite[App. A, TABLE I]{Vitetta_2019} (with $\mathbf{W}%
_{1}\rightarrow \mathbf{W}_{bp,l}$, $\mathbf{W}_{2}\rightarrow \mathbf{W}%
_{pm,l}^{(k)}$, $\mathbf{w}_{1}\rightarrow \mathbf{w}_{bp,l}$ and $\mathbf{w}%
_{2}\rightarrow \mathbf{w}_{pm,l}^{(k)}$). The same considerations apply to
the derivation of the expression (\ref{m_sm_x_l}) of $m_{sm}^{(k)}(\mathbf{x}%
_{l})$.

Step 3) The algorithm for computing $m_{pm,j}^{(k)}(\mathbf{x}_{l}^{(N)})$ (%
\ref{m_pm_x_N_l_j}) \ can be represented as a message passing over the
graphical model shown in Fig. \ref{Fig_6}-b), in which the
pseudo-measurement\ $\mathbf{z}_{l}^{(N)}$ (\ref{z_N_l}) is computed. The
expressions of the involved messages can be derived as follows. Given $%
\overset{\leftarrow }{m}_{be}(\mathbf{x}_{l+1}^{(L)})$ (\ref{m_be_x_L_l+1})
and $m_{sm}^{(k)}(\mathbf{x}_{l}^{(L)})$\ (\ref{m_fe_L_EKF_2}), the message $%
m_{j}^{(k)}(\mathbf{z}_{l}^{(N)})$ can expressed as (see \cite[eqs. (83)-(84)%
]{Vitetta_2018_TF}) 
\begin{equation}
\vec{m}_{j}^{(k)}(\mathbf{z}_{l}^{(N)})=\mathcal{N}\left( \mathbf{z}%
_{l}^{(N)};\mathbf{\check{\eta}}_{z,l,j}^{(k)},\mathbf{\check{C}}%
_{z,l,j}^{(k)}\right) ,  \label{message_z_N}
\end{equation}%
where 
\begin{equation}
\mathbf{\check{\eta}}_{z,l,j}^{(k)}=\mathbf{\tilde{\eta}}_{be,l+1}-\mathbf{A}%
_{l,j}^{(L)}\mathbf{\tilde{\eta}}_{sm,l}^{(k)},  \label{eta_mess_z_N}
\end{equation}%
\begin{equation}
\mathbf{\check{C}}_{z,l,j}^{(k)}=\mathbf{\tilde{C}}_{be,l+1}-\mathbf{A}%
_{l,j}^{(L)}\mathbf{\tilde{C}}_{sm,l}^{(k)}\left( \mathbf{A}%
_{l,j}^{(L)}\right) ^{T}  \label{C_mess_Z_N_bis}
\end{equation}%
and $\mathbf{A}_{l,j}^{(L)}=\mathbf{A}_{l}^{(L)}(\mathbf{x}_{fp,l,j}^{(N)})$%
. Then, $\vec{m}_{j}^{(k)}(\mathbf{z}_{l}^{(N)})$ (\ref{message_z_N}) is
exploited to evaluate (see Fig. \ref{Fig_6}-b)) 
\begin{equation}
\vec{m}_{pm,j}^{(k)}\left( \mathbf{x}_{l}^{(N)}\right) =\int \vec{m}%
_{j}\left( \mathbf{z}_{l}^{(N)}\right) f\left( \mathbf{z}_{l}^{(N)}\left%
\vert \mathbf{x}_{fp,l,j}^{(N)}\right. \right) d\mathbf{z}_{l}^{(N)}.
\label{m_pm_x_N_l_j_first}
\end{equation}%
Substituting eq. (\ref{message_z_N}) and $f(\mathbf{z}_{l}^{(N)}|\mathbf{x}%
_{fp,l,j}^{(N)})=\mathcal{N}(\mathbf{z}_{l}^{(N)};\mathbf{f}_{l,j}^{(L)},%
\mathbf{C}_{w}^{(N)})$ (see eq. (\ref{z_N_l_bis})) in the RHS of the last
expression and evaluating the resulting integral (on the basis of formula
no. 4 of \cite[App. A, TABLE\ II]{Vitetta_2019}) yields eq. (\ref%
{m_pm_x_N_l_j}).

Step 4) The expression (\ref{weight_bpa}) of the weight $w_{bp,l,j}^{(k)}$
is derived as follows. First, we substitute $f(\mathbf{x}_{l+1}^{(N)}/%
\mathbf{x}_{l}^{(N)},\mathbf{x}_{l}^{(L)})=\mathcal{N(}\mathbf{x}%
_{l+1}^{(N)} $; $\mathbf{A}_{l}^{(N)}(\mathbf{x}_{l}^{(N)})\mathbf{x}%
_{l}^{(L)}+\mathbf{f}_{l}^{(N)}(\mathbf{x}_{l}^{(N)}),\mathbf{C}_{w}^{(N)})$
(see eq. (\ref{eq:XL_update}) with $Z=N$), and the expressions of the
messages $\overset{\leftarrow }{m}_{be}(\mathbf{x}_{l+1}^{(N)})$ (\ref%
{mess_be_N_l}) and $m_{sm}^{(k)}(\mathbf{x}_{l}^{(L)})$ (\ref{m_fe_L_EKF_2})
in the RHS\ of eq. (\ref{weight_bp}). Then, the resulting integral is solved
by applying formula no. 1 of \cite[App. A, TABLE\ II]{Vitetta_2019} in the
integration with respect to $\mathbf{x}_{l}^{(L)}$ and the sifting property
of the Dirac delta function in the integration with respect to $\mathbf{x}%
_{l+1}^{(N)}$.

Step 5) The expression (\ref{eq:weight_before_resampling}) of the weight $%
w_{fe1,l,j}^{(k)}$ is derived as follows. First, we substitute $f(\mathbf{y}%
_{l}|\mathbf{x}_{fp,l,j}^{(N)},\,\mathbf{x}_{l}^{(L)})=\mathcal{N(}\mathbf{y}%
_{l}$; $\mathbf{g}_{l,j}+\mathbf{B}_{l,j}\mathbf{x}_{l}^{(L)},\mathbf{C}%
_{e}) $ (with $\mathbf{B}_{l,j}\triangleq \mathbf{B}_{l}(\mathbf{x}%
_{fp,l,j}^{(N)}) $ and $\mathbf{g}_{l,j}\triangleq \mathbf{g}_{l}(\mathbf{x}%
_{fp,l,j}^{(N)})$; see eq. (\ref{eq:y_t})), and eq. (\ref{m_fe_L_EKF_2}) in
eq. (\ref{eq:mess_ms_PF}). Then, the resulting integral is solved by
applying formula no. 1 of \cite[App. A, TABLE\ II]{Vitetta_2019}.

\textbf{Phase III} - The expression (\ref{m_be2_xb}) of $\overset{\leftarrow 
}{m}_{be2,l}\left( \mathbf{x}_{l}\right) $ results from the application of
formula no. 2 of \cite[App. A, TABLE I]{Vitetta_2019} to eq. (\ref{m_be2_x}).

\end{document}